\documentclass[prd,aps,amsmath,amssymb]{revtex4}

\usepackage[english]{babel}
\usepackage[utf8]{inputenc}
\usepackage{graphicx,graphics}
\usepackage{dcolumn}
\usepackage{bm}
\usepackage{hyperref}
\usepackage{blindtext}
\usepackage{mathrsfs}
\usepackage{pstricks}
\usepackage{color}
\usepackage{slashed}
\usepackage{amsmath}
\usepackage{epsfig}
\usepackage{amsfonts}
\usepackage{amssymb}

\def\sgn{\mathrm{sgn}}

\def\beq{\begin{equation}}
\def\eeq{\end{equation}}
\def\bea{\begin{eqnarray}}
\def\eea{\end{eqnarray}}
\def\nn{\nonumber}

\def\Tr{\textrm{Tr}}

\begin{document}

\title{Entanglement dynamics in a Kerr spacetime}
\author{G. Menezes}
\email{gsantosmenez@umass.edu}
\affiliation{Department of Physics, University of Massachusetts, Amherst, Massachusetts 01003, USA}
\affiliation{Departamento de F\'isica, Universidade Fe\-de\-ral Rural do Rio de Janeiro, 23897-000 Serop\'edica, RJ, Brazil.}
%

\begin{abstract}
We consider the entanglement dynamics between two-level atoms in a rotating black-hole background. In our model the two-atom system is envisaged as an open system coupled with a massless scalar field prepared in one of the physical vacuum states of interest. We employ the quantum master equation in the Born-Markov approximation in order to describe the time evolution of the atomic subsystem. We investigate two different states of motion for the atoms, namely static atoms and also stationary atoms with zero angular momentum. The purpose of this work is to expound the impact on the creation of entanglement coming from the combined action of the different physical processes underlying the Hawking effect and the Unruh-Starobinskii effect. We demonstrate that, in the scenario of rotating black holes, the degree of quantum entanglement is significantly modified due to the phenomenon of superradiance in comparison with the analogous cases in a Schwarzschild spacetime. In the perspective of a zero angular momentum observer (ZAMO), one is allowed to probe entanglement dynamics inside the ergosphere, since static observers cannot exist within such a region. On the other hand, the presence of superradiant modes could be a source for violation of complete positivity. This is verified when the quantum field is prepared in the Frolov-Thorne vacuum state. In this exceptional situation, we raise the possibility that the lost of complete positivity is due to the breakdown of the Markovian approximation, which means that any arbitrary physically admissible initial state of the two atoms would not be capable to hold, with time evolution, its interpretation as a physical state inasmuch as negative probabilities are generated by the dynamical map. 
\end{abstract}


\maketitle

\section{Introduction}
\label{intro}

Entanglement and superposition have long been acknowledged as distinguishing features of quantum theory. Entanglement has played a crucial role in several investigations regarding the nature of quantum measurements, most notoriously in the developments of quantum communications~\cite{1,haroche}. In turn, radiative processes of atoms have been shown to be of paramount importance in studies of entanglement decay~\cite{yang,eberly,yun,aga}. Boundary effects are also relevant in this context, since they can induce enhancement or inhibition in the entanglement decay~\cite{jhep}. On the other hand, studies of entanglement dynamics in random media has shown that random fluctuations have the effect of attenuate the nonlocal decoherence~\cite{pra17}. All such investigations suggest that radiative processes of maximally entangled states are to be distinguished from the non-entangled states, as has been emphasized in the literature~\cite{rep}.

The emerging field of relativistic quantum information has been lately drawing increasing attention within the scientific community. Its foundations lie on the combination of methods coming from field theory in curved spacetime and quantum information to study quantum effects induced by gravitational interactions in order to gather information on spacetime. In turn, the subject also addresses questions related with the researches on the role depicted by relativistic settings in quantum information processes. The coupling of atoms with quantum fields is a determinative issue to be considered in order to boost the investigations regarding decoherence properties~\cite{hu3,hu2,hu1}. Some important works formulated within different settings are given by Refs.~\cite{ivet1,ivet2,hu1,cliche2,reznik1,cliche1,martin2}. In turn, a broad compilation of results in this area can be found in the Ref.~\cite{rideout1}. Many of such works provide substantial evidence in favour of the interpretation that entanglement is manifestly an observer-dependent quantity. On the other hand, recent investigations propose that the interplay between vacuum fluctuations and the radiation reaction is an important ingredient in the radiative processes of entangled atoms~\cite{gabriel1,gabriel3,sch}. Concerning this framework, we also refer the reader the Ref.~\cite{prd17}.

In this work we seek to contribute to the foregoing discussion by studying the asymptotic entanglement
of two mutually independent two-level atoms (which we call qubits) outside a rotating black hole. An important concept associated with such objects is the classical phenomenon of wave amplification which came to be known as superradiance~\cite{zeldovich,misner,teu3}. Recent works have been carried out involving black-hole superradiance in different branches of physics~\cite{review}, including connections to dark-matter searches and to physics beyond the Standard Model~\cite{dub,pani,brito}. The study of relativistic jets are also one of the foremost concerns of relevant research regarding astrophysical purposes~\cite{meier}. On the quantum-theory side, the Unruh-Starobinskii process arises as the quantum analog of the superradiance~\cite{sta,unruh-sta,davies}. It predicts the production of particles by the rotational motion of the black hole. This vacuum instability should be distinguished from the Hawking effect~\cite{haw1}. In any case, one could expect that investigations comprising the Unruh-Starobinskii effect are to be pertinent concerning different areas of research, as for instance the studies of emission of jets from accretion disks around supermassive rotating black holes.

In this paper we explore the joint action of Hawking and Unruh-Starobinskii effects by means of the entanglement generation in an open-system framework. We show that a bath of fluctuating vacuum scalar fields outside a Kerr black hole can furnish an indirect coupling between the atoms in order to engender quantum entanglement. In connection with the present studies, we mention that quantum teleportation in the background of Kerr-Newman spacetime was considered in Ref.~\cite{xian2}. In turn, it is known that the Hawking effect of the Kerr spacetime can be understood as the manifestation of thermalization phenomena~\cite{xian}. On the other hand, one can interpret the outcomes to be presented as a nontrivial generalization (in a certain way) of the results derived in Ref.~\cite{hong} which have uncovered the relationship between the Hawking radiation and the spontaneous generation of entanglement between two-level atoms outside a static, spherically symmetric black hole. 

The organization of the paper is as follows. In Sec.~\ref{model} we discuss the coupling of two two-level atoms with quantum massless scalar fields in Kerr spacetime. In Sec.~\ref{master} the quantum master equation is exhibited. This is the device through which one is able to describe the time evolution of the reduced density matrix of the two-qubit system. In Sec.~\ref{ent-static} we examine the asymptotic entanglement between qubits placed at fixed radial distances outside the Kerr black hole. For the sake of clarity we dissect our investigation in four parts, each one related with a physical vacuum of interest, namely the Boulware vacuum state, the Unruh vacuum state and the two vacuum states which are claimed to be equivalent to the Hartle-Hawking vacuum, the Candelas-Chrzanowski-Howard vacuum state and Frolov-Thorne vacuum state. We take the quantum fields as being prepared in each one of such vacua. In Sec.~\ref{ent-stationary} we extend the previous results to encompass the situation of qubits in a stationary motion with a zero angular momentum. Conclusions are given in Sec.~\ref{conclude}. The appendix contains some lengthy derivations of the correlation functions of the scalar field. In this paper we use units such that $\hbar = c = k_B = G = 1$. Summation over repeated indices is assumed, unless otherwise stated.

\section{The coupling of qubits with massless scalar fields in a rotating black-hole spacetime}
\label{model}

Here we are interested in studying the standard approach of open quantum systems regarding quantum entanglement in a particular physical situation. We consider two identical qubits interacting with a quantum massless scalar field outside a Kerr black hole. As well known, the Kerr metric describes the geometry of empty spacetime around a rotating uncharged axially symmetric black hole ~\cite{frolov,ellis}. In Boyer-Lindquist coordinates the Kerr metric reads~\cite{frolov} 
\bea
ds^2 &=& -\left(1 - \frac{2 M r}{\rho^2}\right)\,dt^2 - \frac{4 M a r \sin^2\theta}{\rho^2}\,dt\,d\phi 
+ \frac{\rho^2}{\Delta}\,dr^2 + \rho^2\,d\theta^2 + \frac{\Sigma}{\rho^2}\,\sin^2\theta\,d\phi^2
\nn\\
&=& -\frac{\rho^2\Delta}{\Sigma}\,dt^2 + \frac{\Sigma}{\rho^2}\,\sin^2\theta\,(d\phi - w dt)^2
+ \frac{\rho^2}{\Delta}\,dr^2 + \rho^2\,d\theta^2
\eea
where $w = - g_{0\phi}/g_{00}$ and
\bea
\rho^2 &=& r^2 + a^2\cos^2\theta
\nn\\
\Delta &=& r^2 - 2 M r + a^2
\nn\\
\Sigma &=& (r^2 + a^2)^2 - a^2\Delta\sin^2\theta.
\eea
In this work we use the convention in which the Minkowski metric is given by: $\eta_{\alpha\beta} = 1, \alpha=\beta=1,2,3$, $\eta_{\alpha\beta} = - 1, \alpha=\beta=0$ and $\eta_{\alpha\beta} = 0,\alpha \neq \beta$. 

The Kerr metric is stationary and axially symmetric, with two commuting Killing vectors: $\xi_t$ and $\xi_{\phi}$. The Killing vector $\xi_t$ generates translation in time whereas $\xi_{\phi}$ is a generator of rotations. In order to uniquely specify the vectors $\xi_t$ and $\xi_{\phi}$ one imposes the following conditions: $\xi_{t}$ is the Killing vector which is timelike at infinity with norm $\xi_{t}^2 = -1$; and the integral lines of the Killing vector field $\xi_{\phi}$ are closed. The Kerr metric possesses two parameters, namely the mass $M$ and rotation parameter $a \leq M$; the latter is connected with the angular momentum of the black hole $J$ by $a = J/M$. 

The Killing vector $\xi_t$ becomes spacelike at a region outside the rotating black hole which is called ergosphere. In fact, $\xi_t$ becomes null at the boundary of the ergosphere. This is the static limit surface in which the component $g_{00}$ vanishes. In this region objects must rotate with the black hole. One has that:
\beq
r_{st} = M + \sqrt{M^2 - a^2\cos^2\theta}.
\eeq
Any stationary and axisymmetric spacetime with an event horizon must be endowed with an ergoregion~\cite{cardoso}. In turn, ergoregions can take place in rotating spacetimes with no horizons, as is the case with rapidly rotating neutron stars~\cite{rot1, rot2}. On the other hand, the equation $\Delta = 0$ has two roots:
\beq
r_{\pm} = M \pm \sqrt{M^2 - a^2}.
\eeq
These are coordinate singularities. The surface determined by $r_{+}$ defines the (outer) event horizon. The angular velocity of the event horizon is given by
\beq
\Omega_{H} = \frac{a}{r_{+}^{2} + a^2}.
\label{angular-bh}
\eeq
Let us consider two qubits of vanishing spatial separation following a stationary trajectory in the Kerr spacetime. The stationary trajectory condition guarantees the existence of stationary states. The generic Hamiltonian of the two-qubit system is given by:
\beq
H_A(\tau) = \frac{\omega_0}{2}\,{\bf n}\cdot\biggl[\boldsymbol\sigma_{(1)}(\tau) + \boldsymbol\sigma_{(2)}(\tau)\biggr],
\label{ha-kerrbh}
\eeq
where $\boldsymbol\sigma_{(1)} = \boldsymbol\sigma\otimes\sigma_0$, $\boldsymbol\sigma_{(2)} = \sigma_0\otimes \boldsymbol\sigma$, $\sigma^0$ is the unit $2 \times 2$ unit matrix, $\sigma^j$, $j = 1, 2, 3$ are the Pauli matrices and ${\bf n}$ is a unit vector. Also, $\tau$ is the proper time of the qubits, while $\omega_0$ represents the gap between the two energy eigenvalues. The interaction of the qubits with the external scalar fields is assumed to be weak; it can then be described by an Hamiltonian $H_{I}$ that is linear in both qubits and field variables
\beq
H_{I} = \lambda \sum_{a=1}^{2}\,\sigma^{\mu}_{(a)}\varphi_{\mu}(x(\tau_{a})),
\label{int}
\eeq
with $\mu = 0,\ldots,3$ and $\lambda \ll 1$ is the dimensionless coupling constant between the field and the qubits. Notice that the coupling is effective solely on the trajectory $x(\tau)$ of the qubits. The field operators $\varphi_{\mu}$ represent the external fields and satisfy the massless wave equation in Kerr spacetime. These can be expressed in the standard fashion
\beq
\varphi_{\mu}(x) = \sum_{k=1}^{N}\,\Bigl[\chi_{\mu}^{k}\phi^{-}_{k}(x) + \chi_{\mu}^{k\,*}\phi^{+}_{k}(x)\Bigr],
\eeq
in terms of the positive (negative) energy $\phi^{-}_{k}(x)$ ($\phi^{+}_{k}(x)$) field operators relative to a set of $N$ independent, massless, free scalar fields with a total Hamiltonian $H_{\varphi}(\tau)$. By assuming the field components to be independent,
\beq
\sum_{k=1}^{N}\chi_{\mu}^{k}\chi_{\nu}^{k\,*} = \delta_{\mu\nu}
\eeq
one gets the usual expression for the free scalar field Hamiltonian in terms of creation and annihilation operators. Here we neglect the zero-point energy.

\section{The master equation approach and entanglement}
\label{master}

In this section we discuss the time evolution of the reduced qubit system within the master equation approach. We employ the Born approximation which assumes that initial interaction-induced correlations between the subsystems can be neglected. Taking into account such an approximation, one takes the initial density operator of the whole system as being described by the density operator $\rho(0) = \rho_{A}(0) \otimes \rho_{F}$, where $\rho_{A}(0)$ is the initial density operator associated with the qubit system whereas $\rho_{F}$ is the analogous one for the set of quantum scalar fields, which are assumed to be stationary. The latter is given by $\rho_{F} = |0\rangle\langle 0|$, where $|0\rangle$ is a vacuum state for the quantum fields, to be discussed in due course. For the initial state of the qubits, we consider  a separable state provided by the direct product of two pure states
\beq
\rho_A(0) = \rho_{\nu}\otimes\rho_{\mu},
\label{separable}
\eeq
where
$$
\rho_{\delta} = \frac{1}{2}\left(1 + \boldsymbol\delta\cdot\boldsymbol\sigma\right),
$$
and $\boldsymbol\delta$ stands for the unit vectors $\boldsymbol\mu, \boldsymbol\nu$. Observe that $\rho_A(0)$ has non-negative eigenvalues irrespective of the choice for the unit vectors. We discuss later on some specific choices for the the unit vectors $\boldsymbol\mu$ and $\boldsymbol\nu$. In this case we would be interested to ascertain whether quantum entanglement between the qubits can be generated by the coupling with the quantum fields.

The subdynamics describing the evolution of the subsystem formed by the two qubits is obtained after the procedure of taking the trace over the field degrees of freedom~\cite{edavies1,edavies2,lindblad,koss,cohen4,ben1,ben2,ben3}. Within the Markovian approximation, the evolution of the reduced density operator $\rho_{A}(\tau)$ can be written in the standard Kossakowski-Lindblad form
\beq
\frac{\partial \rho_{A}(\tau)}{\partial \tau} = -i[H_{\textrm{eff}}, \rho_{A}(\tau)] + {\cal L}[\rho_{A}(\tau)],
\label{kerr-master}
\eeq
where
\beq
{\cal L}[\rho_{A}(\tau)] = \frac{1}{2}{\cal K}_{ij}\sum_{a,b = 1}^{2}\,\left\{2\sigma_{(b)}^{j}\,
\rho_{A}(\tau)\sigma_{(a)}^{i} - \{\sigma_{(a)}^{i}\,\sigma_{(b)}^{j},\rho_{A}(\tau)\}\right\},
\label{kerr-nonunitary}
\eeq
and
\beq
H_{\textrm{eff}} = H_A - \frac{i}{2}{\cal H}_{ij}\sum_{a,b=1}^{2}\,\sigma_{(a)}^{i}\sigma_{(b)}^{j}.
\eeq
The so-called Kossakowski matrix $K(\omega_0)$ reads
\beq
K(\omega_0) = \begin{pmatrix}
  {\cal K}(\omega_0) & {\cal K}(\omega_0) \\
   {\cal K}(\omega_0) & {\cal K}(\omega_0)
 \end{pmatrix} 
\label{koss}
\eeq
where the single-qubit Kossakowski matrix ${\cal K}$ is given by
\beq
{\cal K}_{ij}(\omega_0) = A_{+}(\omega_0)\delta_{ij} 
+ {\tilde A}(\omega_0)\,n_{i}n_{j} - i A_{-}(\omega_0)\,\epsilon_{ijm}\,n_{m}.
\label{single-koss}
\eeq
In addition one has that
\beq
{\cal H}_{ij}(\omega_0) = h_{+}(\omega_0)\delta_{ij} 
+ {\tilde h}(\omega_0)\,n_{i}n_{j} - i h_{-}(\omega_0)\,\epsilon_{ijm}\,n_{m},
\eeq
where
\bea
A_{\pm}(\omega_0) &=& \frac{1}{2}\Bigl[{\cal G}(\omega_0) 
\pm {\cal G}(-\omega_0)\Bigr]
\nn\\
{\tilde A}(\omega_0) &=& {\cal G}(0) - A_{+}(\omega_0)
\nn\\
h_{\pm}(\omega_0) &=& \frac{1}{2}\Bigl[{\cal D}(\omega_0) 
\pm {\cal D}(-\omega_0)\Bigr]
\nn\\
{\tilde h}(\omega_0) &=& {\cal D}(0) - h_{+}(\omega_0).
\eea
In the above equations we have introduced the following Fourier and Hilbert transform of the scalar field correlations:
\beq
{\cal G}(\varepsilon\omega_0) = \int_{-\infty}^{\infty}d\Delta\tau\,e^{i\varepsilon\omega_0(\tau - \tau')}
\,D^{+}(x(\tau),x(\tau')),
\label{fourier}
\eeq
and 
\beq
{\cal D}(\varepsilon\omega_0) =  \int_{-\infty}^{\infty}d\Delta\tau\,e^{i\varepsilon\omega_0(\tau-\tau')}\,
\sgn(\tau-\tau')\,D^{+}(x(\tau),x(\tau'))
= \frac{\textrm{P}}{\pi i}\int_{-\infty}^{\infty}d\lambda\,\frac{{\cal G}(\lambda)}{\lambda - \varepsilon\omega_0},
\eeq
where $\Delta\tau = \tau - \tau'$, $\varepsilon = \pm$, $\textrm{P}$ denotes the principal value and $D^{+}(x(\tau),x(\tau'))$ is the standard positive-frequency Wightman function for a massless scalar field. Observe the explicit dependence on the world line $x^{\mu}(\tau)$ of the qubits, as a consequence of Eq.~(\ref{int}). It is here that the entrance of the qubits' worldline in the master equation becomes manifest. In the Appendix we discuss with some detail the different scalar Wightman functions that arise in the context of the Kerr metric. In turn, note that the contributions ${\cal D}(\epsilon\omega_0)$ produce the usual Lamb shift in the energy eigenvalues, as well as a direct two-qubit coupling term. Since we are interested in gravitational-induced effects (in particular the ones possibly related with quantum entanglement), we concentrate on the study of the efects induced by the dissipative part ${\cal L}[\rho_{A}(\tau)]$.

In the investigations concerning open quantum systems, one usually constructs solutions of master equations of so-called Liouville type, $\partial \rho_{A}(\tau)/\partial \tau = \mathbb{D}[\rho_{A}(\tau)]$, where $\mathbb{D}$ is a linear operator defined on the space of reduced density operators $\rho_{A}(\tau)$ representing the two-qubit state at a proper time $\tau$. Memory effects are expected to emerge in this construction, but they are only relevant on short time scales. Consideration of time evolutions on a slow time scale implies in the usage of a suitable Markovian approximation. Such an approximation provides a consistent description of the reduced dynamics of the qubit system, in which one must consider time evolutions comprising one-parameter semigroups of dynamical linear maps 
$\gamma_{\tau} = \exp\{\tau \mathbb{D}\}$. These map positive initial states of the qubits into positive density operators for $\tau > 0$. 

In the case studied in this paper, the Markovian regime consists in considering the dynamics of the reduced qubit system on time scales longer than the decay time of the field correlations. Within such time scales, the dynamics of the qubit system is expected to be unraveled from that of the total system and effectively described by the above master equation, provided that the interaction between the qubits and the quantum fields is sufficiently weak. This is the case assumed in this work. In order to attain a reduced dynamics described by a Markovian prescription, one makes use of the convenient procedure of rescaling the time variable as $\tau \to \tau/\lambda^2$ and then takes the limit $\lambda \to 0$, following the mathematically precise operation of the weak-coupling limit as described in Ref.~\cite{ben3} (see also references cited therein). After this procedure, the evolution equation shall achieve a well-defined limit. 

An important feature related to the above considerations is the notion of {\it complete positivity}. This is particularly important for entangled states, since such states could be mapped out of the physical state space by the time evolution. In this case one must assume that 
$\gamma_{\tau}$ is a completely positive map for all $\tau \geq 0$. This is ensured by a theorem which states that this is only possible if and only if the Kossakowski matrix is positive definite~\cite{ben3}. In the present context, one may employ Sylvester's criterion to prove that a necessary and sufficient for complete positivity of the Kossakowski matrix given by Eq.~(\ref{single-koss}) is that $0 < A_{-} \leq A_{+}$. This is intimately connected with a careful account of the Markovian approximation. This will be further discussed for each of the situations studied in the present work. 

The physical interpretation of complete positiviy can be established as follows. Since the spontaneous excitation rate of a particle detector from the initial ground state $|g\rangle$ to the excited state $|e\rangle$ is given by~\cite{ben2,hong2}
\beq
\Gamma_{|g\rangle \to |e\rangle} = 2 \bigl[A_{+}(\omega_0) - A_{-}(\omega_0)\bigr] = 2 {\cal G}(-\omega_0)
\label{prob}
\eeq
one promptly notices that complete positivity for the time evolution of the atomic subsystem amounts to positivity for spontaneous excitation rates. In turn, since the choice of a physically consistent Markovian prescription must result in semigroups $\gamma_{\tau}$ comprising completely positive maps, one may interpret this result as a consequence of the validity of the Markovian approximation.

It is convenient to write the reduced density operator as
\beq
\rho_{A}(\tau) = \frac{1}{4}\,\rho_{\mu\nu}(\tau)\Sigma_{\mu\nu},
\eeq
where $\Sigma_{\mu\nu} = \sigma_{\mu}\otimes\sigma_{\nu}$ and $\rho_{\mu\nu}(\tau)$ is the Bloch matrix. With the normalization condition $\Tr[\rho_{A}(\tau)] = 1$, and imposing that $\rho_{A}(\tau)$ should be Hermitian, one has that $\rho_{00} = 1$ and $\rho_{i0}(\tau), \rho_{0i}(\tau), \rho_{ij}(\tau)$ are real. Explicitly one has that
\bea
\rho_{A}(\tau) &=&\,\frac{1}{4}\left[{\bf 1}_{4 \times 4} + \rho_{0i}(\tau)\,\sigma_{(2)}^{i}
+ \rho_{i0}(\tau)\,\sigma_{(1)}^{i}+ \rho_{ij}(\tau)\,\sigma_{(1)}^{i}\sigma_{(2)}^{j}\right].
\eea
By disregarding the Hamiltonian contribution in Eq.~(\ref{kerr-master}) and with the elements of the single-qubit Kossakowski matrix as above, as well as using the aforementioned decomposition for $\rho_{A}$, we obtain the following evolution equations for the components of $\rho_{A}(\tau)$:
\beq
\frac{\partial\rho_{0i}}{\partial \tau} = -2\Bigl\{\bigl[(2A_{+} + {\tilde A})\delta_{ik}
- {\tilde A} n_{i}n_{k}\bigr]\rho_{0k}(\tau) - A_{-} n_{k}\rho_{ik}(\tau)\Bigr\} - 2A_{-}(2+\rho^{m}_{\,\,\,m})n_{i}
\eeq
\beq
\frac{\partial\rho_{i0}}{\partial \tau} = -2\Bigl\{\bigl[(2A_{+} + {\tilde A})\delta_{ik}
- {\tilde A} n_{i}n_{k}\bigr]\rho_{k0}(\tau) 
- A_{-} n_{k}\rho_{ki}(\tau)\Bigr\} - 2A_{-}(2+\rho^{m}_{\,\,\,m})n_{i}
\eeq
\bea
\frac{\partial\rho_{ij}}{\partial \tau} &=& - 4\Bigl\{(2A_{+} + {\tilde A})\rho_{ij}(\tau) + (A_{+} + {\tilde A})\rho_{ji}(\tau)
-\bigl[(A_{+}  + {\tilde A})\delta_{ij} - {\tilde A} n_{i}n_{j}\bigr]\rho^{m}_{\,\,\,m}\Bigr\}
\nn\\ 
&-&\, 4 A_{-}[n_{i}\rho_{0j}(\tau) + n_{j}\rho_{i0}(\tau)] - 2A_{-}[n_{i}\rho_{j0}(\tau) + n_{j}\rho_{0i}(\tau)] 
\nn\\
&+&\,2\Bigl\{A_{-}\delta_{ij}n_{k}[\rho_{k0}(\tau)+\rho_{0k}(\tau)] + {\tilde A} n_{i}n_{k}[\rho_{kj}(\tau) + 2\rho_{jk}(\tau)]
+ {\tilde A} n_{j} n_{k}[\rho_{ik}(\tau) + 2\rho_{ki}(\tau)]\Bigr\} 
\nn\\
&-&\, 4{\tilde A}\delta_{ij}n_{k}n_{l}\rho_{kl}(\tau).
\eea
For a detailed discussion on the derivation of such results, see Ref.~\cite{ben3}. Concerning the trace $\rho^{m}_{\,\,\,m}$, this is a constant of motion, as a consequence of the last equation above. Yet, the requirement of positivity of the initial density operator $\rho_{A}(0)$ implies a constraint on the possible values that $\rho^{m}_{\,\,\,m}$ can assume: $-3 \leq \rho^{m}_{\,\,\,m} \leq 1$.

Let us suppose the existence of a equilibrium state $\tilde{\rho}_{A}$ such that ${\cal L}[\tilde{\rho}_{A}(\tau)] = 0$. In this case one obtains~\cite{ben3}
\beq
\tilde{\rho}_{0k} = \tilde{\rho}_{k0} = -\frac{{\cal A} (\rho^{m}_{\,\,\,m} + 3)}{3 + {\cal A}^2}\,n_{k}
\eeq
and
\beq
\tilde{\rho}_{ij} = 
\frac{\left(\rho^{m}_{\,\,\,m} - {\cal A}^2\right)\delta_{ij}}{3+ {\cal A}^2}
+\frac{{\cal A}^2(\rho^{m}_{\,\,\,m} +3) n_{i} n_{j}}{3 + {\cal A}^2} 
\eeq
where ${\cal A}=A_{-}/A_{+} \leq 1$ is positive. Notice that for ${\cal A} = 1$ the equilibrium state only depends on the initial state of the qubits.

In order to verify whether the asymptotic equilibrium state of the qubits is described by an entangled state, one must adopt a convenient criterium as a measure of quantum entanglement. For the case of two two-level systems, this can be suitably provided by the concurrence ${\cal C}[\rho]$, which is a monotonically increasing function of the entanglement of formation. The concurrence acquires a zero value for separable states, whereas for maximally entangled states it is equal to one. In order to compute the concurrence of any $4\times4$ density matrix $\rho$ representing the state of two qubits, one calculates the square roots $\lambda$ of each of the four eigenvalues of the matrix $\rho(\sigma^{2}\otimes\sigma^{2})\rho^{T}(\sigma^{2}\otimes\sigma^{2})$, where $T$ denotes transposition. Then one considers such values in decreasing order: $\lambda_{1} \geq \lambda_{2} \geq \lambda_{3} \geq \lambda_{4}$. The concurrence of $\rho$ is defined to be ${\cal C}[\rho] = \textrm{max}\{\lambda_{1} - \lambda_{2} - \lambda_{3} - \lambda_{4},0\}$. In the case of the asymptotic state derived above, the concurrence has a simple analytical form
\beq
{\cal C}[\tilde{\rho}] = \textrm{max}\left\{\frac{{\cal A}^2 (\rho^{m}_{\,\,\,m} +5)-3 (\rho^{m}_{\,\,\,m} +1)}
{2 \left({\cal A}^2+3\right)},0\right\}.
\eeq
For the case of qubits initially prepared in the separable state~(\ref{separable}), one has that $\rho^{m}_{\,\,\,m} = \boldsymbol\mu \cdot \boldsymbol\nu$. Note that with unit vectors such that $\boldsymbol\mu = \boldsymbol\nu = (0,0,1)$, the asymptotic concurrence is always zero. So in this scenario one is guaranteed not to obtain an asymptotic entangled state for the qubits. In order to reach a nontrivial result, one must then consider a different setup. For instance, one can readily notice that the asymptotic entanglement is maximized for $\boldsymbol\mu = -\boldsymbol\nu$. In this case
\beq
{\cal C}[\tilde{\rho}] = \frac{2{\cal A}^2}{{\cal A}^2+3}.
\label{sep}
\eeq
Notice that one obtains the maximum value ${\cal C}[\tilde{\rho}] = 1/2$ for ${\cal A} = 1$. Expression~(\ref{sep}) is the one we shall systematically employ in the subsequent investigations of quantum entanglement in the Kerr spacetime.

\section{Asymptotic entanglement and complete positivity of open quantum dynamics: the case of static qubits}
\label{ent-static}

The next sections are devoted to the study of the asymptotic equilibrium state as well as the complete positivity of the time evolution of the reduced qubit system in a rotating black-hole background. First let us consider the simple case of static qubits. A static observer in Kerr spacetime (which cannot exist everywhere due to the presence of the ergoregion) is defined as an observer with four velocity given by 
\beq
u^{\mu} = \xi^{\mu}_{t}/|\xi_{t}^2|^{1/2}.
\eeq
The coordinates $r, \theta, \phi$ are constants along its worldline. In this case, the qubits follow the world line given by $x^{\mu}(\tau) = (t(\tau),r,\theta,\phi)$, where $t(\tau) = \int d\tau u^0 = \int d\tau |g_{00}|^{-1/2} = \tau |g_{00}|^{-1/2}$, with $g_{00}(r,\theta) = -(1 - 2 M r/\rho^2)$. In this situation the static qubits remain in the region $r > r_{st}$. In turn, as mentioned above, we consider the scalar fields prepared in each of the physical vacuum states of interest, namely the Boulware vacuum state, the Unruh vacuum state, the Candelas-Chrzanowski-Howard vacuum state and Frolov-Thorne vacuum state. For further discussions on such vacuum states, we refer the reader the Ref.~\cite{ote}. In any case, the Appendix contains some brief discussions on such states as well as calculations of the associated scalar Wightman functions.

\subsection{The Boulware vacuum state}

In Schwarzschild spacetime, the Boulware vacuum is the proper choice of vacuum state for quantum fields in the vicinity of an isolated, cold neutron star~\cite{boul}. In Kerr spacetime, the existence of superradiant modes renders the discussion more involved. Two kinds of Boulware vacuum states can be defined~\cite{ote}. Such states do not strictly agree with the idea of a Boulware state in Schwarzschild spacetime as the most empty state at infinity; this is a result of the Unruh-Starobinskii effect~\cite{unruh-sta}. The non-existence of a Boulware vacuum state (as understood in Schwarzschild spacetime) is closely connected with the fact that a true Hartle-Hawking state cannot be obtained in Kerr spacetime~\cite{wald}. 

Let us calculate the Fourier transform of the positive-frequency scalar Wightman function with respect to the Boulware vacuum state. Taking into account the results presented in the Appendix, expression~(\ref{fourier}) can be written as (the subindex in ${\cal G}$ denotes the vacuum state just considered)
\bea
{\cal G}_{B}(\varepsilon\omega_0) &=& \frac{1}{8\pi^2}\sum_{l,m}\,
\,\int_{-\infty}^{\infty}du\,\Biggl[\int_{0}^{\infty}\,d\omega\,{\cal P}^{+}_{\omega l m}(r, \theta)\, 
(\cos(w u) - i \sin(w u)) 
\nn\\
&+&\, \int_{0}^{\infty}\,d\bar{\omega} {\cal P}^{-}_{\omega l m}(r, \theta)\,
(\cos(w u) - i \sin(w u)) \Biggr]\,e^{i \varepsilon \omega_0 u},
\label{vf-boul}
\eea
where $u = \Delta\tau = \tau - \tau'$, $w = w(\omega) = \omega |g_{00}|^{-1/2}$ and
\bea
{\cal P}^{\pm}_{\omega l m}(r, \theta) = \frac{|S_{\omega l m}(\cos\theta)|^2\,
|R^{\pm}_{\omega l m} (r)|^2}
{\omega^{\pm}(r^2 + a^2)},
\eea
with $\omega^{+} = \omega$ and $\omega^{-} = \bar{\omega} = \omega - m\Omega_{H}$. All the above quantities are defined in the Appendix. We considered only the past modes in order to derive expression~(\ref{vf-boul}) but a similar analysis can be carried out for the future modes. Actually, one can use the asymptotic forms of the modes presented in the Ref.~\cite{ote} in order to discuss the behavior of ${\cal P}^{\pm}$ near the event horizon and also far away from the black hole. One gets:
\begin{eqnarray}
{\cal P}^{+}_{\omega l m}(r, \theta) &\sim& \frac{|S_{\omega l m}(\cos\theta)|^2}
{\omega}\,
	\begin{cases}
               0 \hspace{90pt} \textrm{at ${\cal H}^{-}$}\\
               (r^2 + a^2)^{-1} \hspace{45pt} \textrm{at ${\cal J}^{-}$}\\
               (r_{+}^2 + a^2)^{-1}\,|{\cal T}^{+}_{\omega l m}|^2 \hspace{11pt} \textrm{at ${\cal H}^{+}$}\\
               (r^2 + a^2)^{-1}\,|{\cal R}^{+}_{\omega l m}|^2 \hspace{10pt} \textrm{at ${\cal J}^{+}$} 
	\end{cases}
\label{pasymp1}
\end{eqnarray}
and
\begin{eqnarray}
{\cal P}^{-}_{\omega l m}(r, \theta) &\sim& \frac{|S_{\omega l m}(\cos\theta)|^2}
{(\omega - m\Omega_{H})}\,
	\begin{cases}
               (r_{+}^2 + a^2)^{-1} \hspace{42pt} \textrm{at ${\cal H}^{-}$}\\
               0 \hspace{89pt} \textrm{at ${\cal J}^{-}$}\\
              (r_{+}^2 + a^2)^{-1}\, |{\cal R}^{-}_{\omega l m}|^2 \hspace{7pt} \textrm{at ${\cal H}^{+}$}\\
               (r^2 + a^2)^{-1}\,|{\cal T}^{-}_{\omega l m}|^2 \hspace{12pt} \textrm{at ${\cal J}^{+}$}, 
	\end{cases}
\label{pasymp2}
\end{eqnarray}
where ${\cal R}, {\cal T}$ are the reflection and transmission coefficients, respectively. 

The time integrals appearing in Eq.~(\ref{vf-boul}) can be solved in a straightforward way and one finds
\bea
{\cal G}_{B}(\varepsilon\omega_0) &=& \frac{|g_{00}|^{1/2}}{4\pi}
\,\sum_{l,m}\Biggl[{\cal P}^{+}_{(\varepsilon\widetilde{\omega}_0) l m}(r, \theta)\theta(\epsilon) 
+ {\cal P}^{-}_{(\varepsilon\widetilde{\omega}_0) l m}(r, \theta)
\theta(\varepsilon \widetilde{\omega}_0 - m\Omega_{H})\Biggr]
\eea
where $\omega_0\,(u^{0})^{-1} = \widetilde{\omega}_0$ and we used that $\delta(ax) = \delta(x)/|a|$. In this way one has
\beq
{\cal A}_{B}(\omega_0) = \frac{{\cal G}_{B}(\omega_0) - {\cal G}_{B}(-\omega_0)}
{{\cal G}_{B}(\omega_0) + {\cal G}_{B}(-\omega_0)} 
= \sum_{l,m} {\cal B}_{lm}(\omega_0)
\eeq
where
\bea
{\cal B}_{lm}(\omega_0) = \frac{{\cal W}_{B}^{lm\,-}(\omega_0)}
{\sum_{l,m}{\cal W}_{B}^{lm\,+}(\omega_0)}
\eea
with
\bea
\widehat{{\cal W}}_{B}^{lm\,\pm}(\omega_0) &=& 
{\cal P}^{+}_{\widetilde{\omega}_0 l m}(r, \theta)
+ {\cal P}^{-}_{\widetilde{\omega}_0 l m}(r, \theta)\,
\theta[\widetilde{\omega}_0 - m\Omega_{H}]
\pm{\cal P}^{-}_{-\widetilde{\omega}_0 l m}(r, \theta)\,
\theta[-\widetilde{\omega}_0 - m\Omega_{H}].
\eea
Since ${\cal G}_{B}(-\omega_0) \neq 0$ one sees from Eq.~(\ref{sep}) that, for qubits initially prepared in the separable state~(\ref{separable}) with $\boldsymbol\mu = -\boldsymbol\nu$, the concurrence reaches its maximum value only at ${\cal J}^{-}$ where ${\cal P}^{-} = 0$. On the other hand, 
at ${\cal J}^{+}$ the asymptotic entanglement does not reach its possible maximum value. 
This can be seen as the result of the fact that the past Boulware vacuum contains at future null infinity an outward flux of particles owing to the Unruh-Starobinskii radiation~\cite{ote}. In addition, in order to assure complete positivity for the evolution of the reduced system, one must assume that ${\cal A}_{B}(\omega_0) \leq 1$, which implies that ${\cal G}_{B}(-\omega_0) \geq 0$. This is clearly satisfied, which implies that the Markovian approximation is well-founded in this case.

\subsection{The Unruh vacuum state}

In Schwarzschild spacetime, the Unruh vacuum state is the appropriate vacuum state which is most important to the gravitational collapse of a massive spherically symmetric body~\cite{unruh}. At spatial infinity this state amounts to an outgoing flux of black-body radiation at the black-hole temperature~\cite{sciama}. In Kerr spacetime one defines two types of Unruh vacuum states: one has a past (future) Unruh state as a state empty at ${\cal J}^{-}$ (${\cal J}^{+}$) but with modes on ${\cal H}^{-}$ (${\cal H}^{+}$) thermally populated. It is the past Unruh state that mimics the state emerging at late times from the collapse of a star to form a black hole. Accordingly, this is the one considered in this work. 

Let us proceed to calculate the Fourier transform of the positive-frequency scalar Wightman function with respect to the (past) Unruh vacuum state. Recalling the results presented in the Appendix, expression~(\ref{fourier}) becomes
\bea
{\cal G}_{U}(\varepsilon\omega_0) &=& \frac{1}{8\pi^2}\sum_{l,m}\,
\,\int_{-\infty}^{\infty}du\,\Biggl\{\int_{0}^{\infty}\,d\omega\,{\cal P}^{+}_{\omega l m}(r, \theta)\, 
(\cos(w u) - i \sin(w u)) 
\nn\\
&+&\, \int_{0}^{\infty}\,d\bar{\omega} {\cal P}^{-}_{\omega l m}(r, \theta)
\left[\coth\left( \frac{\pi\bar{\omega}}{\kappa_{+}} \right)\cos(w u) - i \sin(w u)\right] \Biggr\}\,
e^{i \varepsilon \omega_0 u},
\eea
where $\kappa_{+}$ is the surface gravity on the outer horizon, given by Eq.~(\ref{surface-gravity}). One can easily solve the time integrals to obtain
\bea
{\cal G}_{U}(\varepsilon\omega_0) &=& \frac{|g_{00}|^{1/2}}{4\pi}
\,\sum_{l,m}\Biggl\{{\cal P}^{+}_{(\varepsilon\widetilde{\omega}_0) l m}(r, \theta)\,\theta(\epsilon)
+ {\cal P}^{-}_{(\varepsilon\widetilde{\omega}_0) l m}(r, \theta)
\left(1 + \frac{1}{e^{\frac{2\pi}{\kappa_{+}}(\varepsilon\widetilde{\omega}_0 - m\Omega_{H})} - 1}\right)
\theta(-m\Omega_{H} + \varepsilon\widetilde{\omega}_0) 
\nn\\
&+&\,\frac{{\cal P}^{-}_{(-\varepsilon\widetilde{\omega}_0) l m}(r, \theta)}{e^{\frac{2\pi}{\kappa_{+}}](-\varepsilon\widetilde{\omega}_0 - m\Omega_{H})} - 1}
\theta(-m\Omega_{H} - \varepsilon\widetilde{\omega}_0)\Biggr\}.
\eea
Note that the Bose-Einstein factor gets modified by a grey-body factor; this means that thermal radiation from the black hole is backscattered off the spacetime curvature. One finds
\beq
{\cal A}_{U}(\omega_0) = \frac{{\cal G}_{U}(\omega_0) - {\cal G}_{U}(-\omega_0)}
{{\cal G}_{U}(\omega_0) + {\cal G}_{U}(-\omega_0)} = \sum_{l,m} {\cal U}_{lm}(\omega_0)
\eeq
where
\bea
{\cal U}_{lm}(\omega_0) = \frac{{\cal W}_{U}^{lm\,-}(\omega_0)}{\sum_{l,m}{\cal W}_{U}^{lm\,+}(\omega_0)}
\eea
with
\bea
{\cal W}_{U}^{lm\,\pm}(\omega_0) &=& 
{\cal P}^{+}_{\widetilde{\omega}_0 l m}(r, \theta)\,
+ {\cal P}^{-}_{\widetilde{\omega}_0 l m}(r, \theta)
\left(1 + \frac{1}{e^{\frac{2\pi}{\kappa_{+}}(\widetilde{\omega}_0 - m\Omega_{H})} - 1}\right)
\theta(-m\Omega_{H} + \widetilde{\omega}_0) 
+\frac{{\cal P}^{-}_{-\widetilde{\omega}_0 l m}(r, \theta)\,\theta(-m\Omega_{H} -\widetilde{\omega}_0)}
{e^{\frac{2\pi}{\kappa_{+}}(-\widetilde{\omega}_0 - m\Omega_{H})} - 1}
\nn\\
&\pm& {\cal P}^{-}_{-\widetilde{\omega}_0 l m}(r, \theta)
\left(1 + \frac{1}{e^{\frac{2\pi}{\kappa_{+}}(-\widetilde{\omega}_0 - m\Omega_{H})} - 1}\right)
\theta(-m\Omega_{H} - \widetilde{\omega}_0) 
\pm\frac{{\cal P}^{-}_{\widetilde{\omega}_0 l m}(r, \theta)}
{e^{\frac{2\pi}{\kappa_{+}}(\widetilde{\omega}_0 - m\Omega_{H})} - 1}
\theta(-m\Omega_{H} +\widetilde{\omega}_0).
\eea
Notice that the concurrence~(\ref{sep}) cannot reach its maximum for a finite $r$. In the present case, this is a consequence of the interplay between superradiance and the Hawking effect. The temperature of the thermal radiation is given by
\beq
T = \frac{\kappa_{+}}{2\pi}\,|g_{00}|^{-1/2},
\label{temp}
\eeq
with $\kappa_{+}/2\pi$ being the usual Hawking temperature of the black hole. As expected, the rotation of the black hole enters into the thermal spectrum as a chemical potential~\cite{birrel}. As for the complete positivity, one must assume that ${\cal A}_{U}(\omega_0) \leq 1$, which is equivalent to 
${\cal G}_{U}(-\omega_0) \geq 0$. In the present case, such a condition is always fulfilled. The Markovian description in this case is thus completely justified. Finally, once again Eqs.~(\ref{pasymp1}) and~(\ref{pasymp2}) reveal that ${\cal A}_{U} \to 1$ in the region ${\cal J}^{-}$. In this region the asymptotic entanglement is maximized.

\subsection{The Candelas-Chrzanowski-Howard vacuum state}

There does not exist an everywhere regular Hadamard state in Kerr spacetime~\cite{wald}. In this scenario, there are attempts in the literature purported to define a thermal state with several properties pertained to the Hartle-Hawking state. In this Subsection we discuss the results associated with the state introduced by Candelas, Chrzanowski and Howard~\cite{candelas2}, which is obtained by thermalizing the in and up modes with respect to their natural energy (for a definition of such modes, see the Appendix or Ref.~\cite{ote}). Such a vacuum state could be described as a past Hartle-Hawking vacuum. Yet, such a state does not respect the simultaneous $t-\phi$ reversal invariance of Kerr spacetime. 

Let us calculate the Fourier transform of the positive-frequency scalar Wightman function concerning the Candelas-Chrzanowski-Howard vacuum state. From the results in the Appendix and~(\ref{fourier}), one gets
\bea
{\cal G}_{CCH}(\varepsilon\omega_0) &=& \frac{1}{8\pi^2}\sum_{l,m}\,
\int_{-\infty}^{\infty}du\,\Biggl\{\int_{0}^{\infty}\,d\omega\,
{\cal P}^{+}_{\omega l m}(r, \theta) 
\left[\coth\left( \frac{\pi\omega}{\kappa_{+}} \right)\cos(w u) - i \sin(w u)\right]
\nn\\
&+&\, \int_{0}^{\infty}\,d\bar{\omega} {\cal P}^{-}_{\omega l m}(r, \theta)
\left[\coth\left( \frac{\pi\bar{\omega}}{\kappa_{+}} \right)\cos(w u) - i \sin(w u)\right]\Biggr\}\,
e^{i \varepsilon \omega_0 u}.
\eea
Similarly as the cases discussed previously, the time integrals can be easily solved and one finds that
\bea
{\cal G}_{CCH}(\varepsilon\omega_0) &=& \frac{|g_{00}|^{1/2}}{4\pi}
\sum_{l,m}\Biggl\{\left[{\cal P}^{+}_{(\varepsilon\widetilde{\omega}_0) l m}(r, \theta)
\left(1 + \frac{1}{e^{\frac{2\pi}{\kappa_{+}}\varepsilon\widetilde{\omega}_0} - 1}\right)\theta(\varepsilon)
+\frac{{\cal P}^{+}_{(|\varepsilon|\widetilde{\omega}_0) l m}(r, \theta)}
{e^{\frac{2\pi}{\kappa_{+}}|\varepsilon|\widetilde{\omega}_0} - 1}\theta(-\varepsilon)
\right.
\nn\\
&+&\,\left. {\cal P}^{-}_{(\varepsilon\widetilde{\omega}_0) l m}(r, \theta)
\left(1 + \frac{1}{e^{\frac{2\pi}{\kappa_{+}}(\varepsilon\widetilde{\omega}_0 - m\Omega_{H})} - 1}\right)
\theta(-m\Omega_{H} + \varepsilon\widetilde{\omega}_0) 
\right.
\nn\\
&+&\,\left. \frac{{\cal P}^{-}_{(-\varepsilon\widetilde{\omega}_0) l m}(r, \theta)}{e^{\frac{2\pi}{\kappa_{+}}](-\varepsilon\widetilde{\omega}_0 - m\Omega_{H})} - 1}
\theta(-m\Omega_{H} - \varepsilon\widetilde{\omega}_0)\right\}.
\label{Fourier-CCH}
\eea
The presence of the Planckian factor in both ${\cal P}^{\pm}$ suggests the existence of thermal radiation outgoing from the horizon and that incoming from infinity. Both are modified by the grey-body factors ${\cal P}^{\pm}$ because of backscattering off the spacetime curvature. In turn, one gets
\beq
{\cal A}_{CCH}(\omega_0) = \frac{{\cal G}_{CCH}(\omega_0) - {\cal G}_{CCH}(-\omega_0)}
{{\cal G}_{CCH}(\omega_0) + {\cal G}_{CCH}(-\omega_0)} = \sum_{l,m} {\cal F}_{lm}(\omega_0)
\eeq
where
\bea
{\cal F}_{lm}(\omega_0) = \frac{{\cal W}_{CCH}^{lm\,-}(\omega_0)}{\sum_{l,m}{\cal W}_{CCH}^{lm\,+}(\omega_0)}
\eea
with
\bea
{\cal W}_{CCH}^{lm\,\pm}(\omega_0) &=& 
{\cal P}^{+}_{\widetilde{\omega}_0 l m}(r, \theta)
\left(1 + \frac{1}{e^{\frac{2\pi}{\kappa_{+}}\widetilde{\omega}_0} - 1}\right)
+ {\cal P}^{-}_{\widetilde{\omega}_0 l m}(r, \theta)
\left(1 + \frac{1}{e^{\frac{2\pi}{\kappa_{+}}(\widetilde{\omega}_0 - m\Omega_{H})} - 1}\right)
\theta(-m\Omega_{H} + \widetilde{\omega}_0) 
\nn\\
&+&\frac{{\cal P}^{-}_{-\widetilde{\omega}_0 l m}(r, \theta)}
{e^{\frac{2\pi}{\kappa_{+}}(-\widetilde{\omega}_0 - m\Omega_{H})} - 1}
\theta(-m\Omega_{H} -\widetilde{\omega}_0)
\pm \frac{{\cal P}^{+}_{\widetilde{\omega}_0 l m}(r, \theta)}
{e^{\frac{2\pi}{\kappa_{+}}\widetilde{\omega}_0} - 1}
\nn\\
&\pm& {\cal P}^{-}_{-\widetilde{\omega}_0 l m}(r, \theta)
\left(1 + \frac{1}{e^{\frac{2\pi}{\kappa_{+}}(-\widetilde{\omega}_0 - m\Omega_{H})} - 1}\right)
\theta(-m\Omega_{H} - \widetilde{\omega}_0) 
\pm\frac{{\cal P}^{-}_{\widetilde{\omega}_0 l m}(r, \theta)}
{e^{\frac{2\pi}{\kappa_{+}}(\widetilde{\omega}_0 - m\Omega_{H})} - 1}
\theta(-m\Omega_{H} +\widetilde{\omega}_0).
\eea
Again the combined effort coming from the phenomena of superradiance and the Hawking effect prevents the concurrence~(\ref{sep}) from achieving its maximum value of $1/2$, even at the region 
${\cal J}^{-}$. This is due to the thermal nature of the Candelas-Chrzanowski-Howard vacuum state. Moreover, if the function $|R^{-}_{\widetilde{\omega} l m}|^2$ can be envisaged as a continuous function of $\widetilde{\omega}$ and remains bounded as $\widetilde{\omega} \to 0$ (which the radial equation~(\ref{radial}) in the Appendix and the analysis in Ref.~\cite{Candelas:80} as $r \to r_{+}$ seem to imply), then ${\cal A}_{CCH}(\omega_0) \to 0$ as the qubits approach the ergosphere. This suggests that the asymptotic state of the atomic subsystem cannot be described by an entangled state for distances arbitrarily close to the ergosphere. In turn, the condition ${\cal A}_{CCH}(\omega_0) \leq 1$, or ${\cal G}_{CCH}(-\omega_0) \geq 0$, ensuring the complete positivity for the evolution of the reduced system, is always satisfied. In other words, the Markovian description is utterly valid for the Candelas-Chrzanowski-Howard vacuum state.

\subsection{The Frolov-Thorne vacuum state}

The second state we consider in this work which comprises some features of the Hartle-Hawking state is the one introduced by
Frolov and Thorne~\cite{fro3} by using the alternative ``$\eta$ formalism" in order to cope with the quantization of the superradiant modes. This state is invariant under simultaneous $t-\phi$ reversal~\cite{ote}. In turn, as demonstrated in Ref.~\cite{prd17}, the Frolov-Thorne vacuum state can describe black-hole superradiance in the quantum regime.

Let us calculate the Fourier transform of the positive-frequency scalar Wightman function in the Frolov-Thorne vacuum state. Use of the results in the Appendix together with Eq.~(\ref{fourier}) leads one to
\bea
{\cal G}_{FT}(\varepsilon\omega_0) &=&  \frac{1}{8\pi^2}\sum_{l,m}\,
\int_{-\infty}^{\infty}du\,\Biggl\{\int_{0}^{\infty}\,d\omega\,
{\cal P}^{+}_{\omega l m}(r, \theta) 
\left[\coth\left( \frac{\pi\bar{\omega}}{\kappa_{+}} \right)\cos(w u) - i \sin(w u)\right]
\nn\\
&+&\, \int_{0}^{\infty}\,d\bar{\omega} {\cal P}^{-}_{\omega l m}(r, \theta)
\left[\coth\left( \frac{\pi\bar{\omega}}{\kappa_{+}} \right)\cos(w u) - i \sin(w u)\right]\Biggr\}\,
e^{i \varepsilon \omega_0 u}.
\eea
By solving the time integrals, one has that
\bea
{\cal G}_{FT}(\varepsilon\omega_0) &=& \frac{|g_{00}|^{1/2}}{4\pi}
\sum_{l,m}\Biggl\{\left[{\cal P}^{+}_{(\varepsilon\widetilde{\omega}_0) l m}(r, \theta)
\left(1 + \frac{1}{e^{\frac{2\pi}{\kappa_{+}}(\varepsilon\widetilde{\omega}_0- m\Omega_H)} - 1}\right)\theta(\varepsilon)
+\frac{{\cal P}^{+}_{(|\varepsilon|\widetilde{\omega}_0) l m}(r, \theta)}
{e^{\frac{2\pi}{\kappa_{+}}(|\varepsilon|\widetilde{\omega}_0 -m\Omega_H)} - 1}\theta(-\varepsilon)
\right.
\nn\\
&+&\,\left. {\cal P}^{-}_{(\varepsilon\widetilde{\omega}_0) l m}(r, \theta)
\left(1 + \frac{1}{e^{\frac{2\pi}{\kappa_{+}}(\varepsilon\widetilde{\omega}_0 - m\Omega_{H})} - 1}\right)
\theta(-m\Omega_{H} + \varepsilon\widetilde{\omega}_0) 
\right.
\nn\\
&+&\,\left. \frac{{\cal P}^{-}_{(-\varepsilon\widetilde{\omega}_0) l m}(r, \theta)}{e^{\frac{2\pi}{\kappa_{+}}](-\varepsilon\widetilde{\omega}_0 - m\Omega_{H})} - 1}
\theta(-m\Omega_{H} - \varepsilon\widetilde{\omega}_0)\right\}.
\label{Fourier-FT}
\eea
Notice the difference between Eqs.~(\ref{Fourier-CCH}) and~(\ref{Fourier-FT}); the angular velocity of the black hole enters in all thermal contributions. One gets
\beq
{\cal A}_{FT}(\omega_0) = \frac{{\cal G}_{FT}(\omega_0) - {\cal G}_{FT}(-\omega_0)}
{{\cal G}_{FT}(\omega_0) + {\cal G}_{FT}(-\omega_0)} = \sum_{l,m} {\cal X}_{lm}(\omega_0)
\eeq
where
\bea
{\cal X}_{lm}(\omega_0) = \frac{{\cal W}_{FT}^{lm\,-}(\omega_0)}{\sum_{l,m}{\cal W}_{FT}^{lm\,+}(\omega_0)}
\eea
with
\bea
{\cal W}_{FT}^{lm\,\pm}(\omega_0) &=& 
{\cal P}^{+}_{\widetilde{\omega}_0 l m}(r, \theta)
\left(1 + \frac{1}{e^{\frac{2\pi}{\kappa_{+}}(\widetilde{\omega}_0-m\Omega_H)} - 1}\right)
+ {\cal P}^{-}_{\widetilde{\omega}_0 l m}(r, \theta)
\left(1 + \frac{1}{e^{\frac{2\pi}{\kappa_{+}}(\widetilde{\omega}_0 - m\Omega_{H})} - 1}\right)
\theta(-m\Omega_{H} + \widetilde{\omega}_0) 
\nn\\
&+&\frac{{\cal P}^{-}_{-\widetilde{\omega}_0 l m}(r, \theta)}
{e^{\frac{2\pi}{\kappa_{+}}(-\widetilde{\omega}_0 - m\Omega_{H})} - 1}
\theta(-m\Omega_{H} -\widetilde{\omega}_0)
\pm \frac{{\cal P}^{+}_{\widetilde{\omega}_0 l m}(r, \theta)}
{e^{\frac{2\pi}{\kappa_{+}}(\widetilde{\omega}_0 -m\Omega_H)} - 1}
\nn\\
&\pm& {\cal P}^{-}_{-\widetilde{\omega}_0 l m}(r, \theta)
\left(1 + \frac{1}{e^{\frac{2\pi}{\kappa_{+}}(-\widetilde{\omega}_0 - m\Omega_{H})} - 1}\right)
\theta(-m\Omega_{H} - \widetilde{\omega}_0) 
\pm\frac{{\cal P}^{-}_{\widetilde{\omega}_0 l m}(r, \theta)}
{e^{\frac{2\pi}{\kappa_{+}}(\widetilde{\omega}_0 - m\Omega_{H})} - 1}
\theta(-m\Omega_{H} +\widetilde{\omega}_0).
\eea
As in the previous case, note the emergence of thermal contributions in all terms of the above equations. In turn, again the concurrence~(\ref{sep}) does not reach its maximum value of $1/2$, even at the region ${\cal J}^{-}$. Similarly to the previous case, this is the result of the thermal nature of the Frolov-Thorne vacuum state. On the other hand, the conditions ${\cal A}_{FT}(\omega_0) \leq 1$ or ${\cal G}_{FT}(-\omega_0) \geq 0$ lead us to
\bea
&&\sum_{l,m}
\Biggl[\frac{{\cal P}^{+}_{\widetilde{\omega}_0 l m}(r, \theta)}
{e^{\frac{2\pi}{\kappa_{+}}(\widetilde{\omega}_0 - m\Omega_{H})} - 1}
+ \frac{{\cal P}^{-}_{\widetilde{\omega}_0 l m}(r, \theta)}
{e^{\frac{2\pi}{\kappa_{+}}(\widetilde{\omega}_0 - m\Omega_{H})} - 1}
\theta(-m\Omega_{H} + \widetilde{\omega}_0) 
\nn\\
&+&\,
{\cal P}^{-}_{-\widetilde{\omega}_0 l m}(r, \theta)
\left(1 + \frac{1}{e^{\frac{2\pi}{\kappa_{+}}(-\widetilde{\omega}_0 - m\Omega_{H})} - 1}\right)
\theta(-m\Omega_{H} - \widetilde{\omega}_0)\Biggr] \geq 0.
\eea
As one can easily notice, such a requisite could be violated at the region ${\cal J}^{-}$ as a consequence of the superradiant condition $\widetilde{\omega}_0 < m\Omega_{H}$, from a certain value of $m$. Recalling Eq.~(\ref{prob}) and the interpretation given in Ref.~\cite{prd17}, one can easily understand the origin of such an adversity: This is clearly a consequence of superradiance. In this case the black hole induces stimulated emission. The absorption probability of the black hole in this case is negative, which leads to a negative spontaneous excitation rate. In quantum terminology, this is the result of the Unruh-Starobinskii effect. As will be demonstrated subsequently, this also takes place when one considers the Frolov-Thorne vacuum state from the point of view of stationary qubits with zero angular momentum. In any case one may say that, in the case of the Frolov-Thorne vacuum, the Unruh-Starobinskii effect hinders the Markovian description of the time evolution of the qubit system; as a result, the breakdown of the Markovian approximation spoils the complete positivity for dynamical maps.   

\section{Asymptotic entanglement and complete positivity of open quantum dynamics: the case of stationary qubits with zero angular momentum}
\label{ent-stationary}

In the present section we turn our attentions to the reduced dynamics of the two-qubit system for the case of stationary trajectories, but not necessarily with fixed spatial coordinates. As well known, a stationary observer in Kerr spacetime is an observer whose four-velocity is a combination of the two Killing vectors of the Kerr metric
\beq
u^{\mu} = (\xi^{\mu}_{t} + \Omega\,\xi^{\mu}_{\phi})/|\xi^{\mu}_{t} + \Omega\,\xi^{\mu}_{\phi}| = u^{0}(1,0,0,\Omega)
\eeq
where
$$
|\xi^{\mu}_{t} + \Omega\,\xi^{\mu}_{\phi}|^2 = - g_{\mu\nu}(\xi^{\mu}_{t} + \Omega\,\xi^{\mu}_{\phi})
(\xi^{\nu}_{t} + \Omega\,\xi^{\nu}_{\phi}),
$$
and the angular velocity of the observer is $\Omega = d\phi/dt = u^{\phi}/u^{0}$. The vector $\xi^{\mu}_{t} + \Omega\,\xi^{\mu}_{\phi}$ becomes null at $r = r_{+}$ and stationary observers cannot exist inside this surface, which one identifies with the black hole's event horizon. The quantity
\beq
\Omega_{H} = \frac{a}{r_{+}^{2} + a^2} =  \frac{2 M a r_{+}}{(r_{+}^2 + a^2)^2}
\eeq
is the angular velocity of the black hole. Stationary observers just outside the horizon have an angular velocity equal to $\Omega_{H}$. 

In the general stationary case, the qubits follow the world line given by $x^{\mu}(\tau) = (t(\tau),r,\theta,\phi(\tau))$, where $r, \theta$ are constants and $\phi(\tau) = \Omega t(\tau)$, $t(\tau) = u^{0}\tau$. Here we shall consider the special case in which the qubits have zero angular momentum. This implies that at $r \to \infty$, where the metric becomes flat, one gets $u^{\phi} = 0$. On the other hand, $u^{\phi} = g^{\phi 0}\,u_{0} \neq 0$ for finite $r$. The trajectory of the ZAMO has a non-zero angular velocity:
\bea
\Omega &=& - \frac{g_{\phi 0}}{g_{\phi\phi}} =  \frac{2 M a r}{(r^2 + a^2)^2 - a^2\Delta\sin^2\theta}.
\eea
Notice that $\Omega \leq \Omega_{H}$. The four velocity is given by
\beq 
u_{\mu} = -\alpha\,\delta_{\mu}^{0},\,\,\,u^{\mu} = \alpha^{-1}(1,0,0,\Omega),\,\,\,
\alpha = \sqrt{\frac{\rho^2 \Delta}{\Sigma}}.
\eeq

\subsection{The Boulware vacuum state}

As above, we start our discussion with the Boulware vacuum states. The Fourier transform of the Wightman function is given by
\bea
{\cal G}_{B}(\varepsilon\omega_0) &=& \frac{1}{8\pi^2}\sum_{l,m}\,
\,\int_{-\infty}^{\infty}du\,\Biggl\{\int_{0}^{\infty}\,d\omega\,{\cal P}^{+}_{\omega l m}(r, \theta)\, 
\left[\cos \Bigl((\widetilde{w}-\widetilde{m})u\Bigr) -i\sin \Bigl((\widetilde{w}-\widetilde{m})u\Bigr)\right]
\nn\\
&+&\, \int_{0}^{\infty}\,d\bar{\omega} {\cal P}^{-}_{\omega l m}(r, \theta)\,
\left[\cos \Bigl((\widetilde{w}-\widetilde{m})u\Bigr) -i\sin \Bigl((\widetilde{w}-\widetilde{m})u\Bigr)\right] \Biggr\}\,e^{i \varepsilon \omega_0 u},
\eea
where $\widetilde{w} = \widetilde{w}(\omega) = \omega u^{0}$ and $\widetilde{m} = m\Omega u^{0}$. As previously, such an expression was derived for the past modes. Solving the time integrals one gets (recall that $\Omega \leq \Omega_{H}$)
\bea
{\cal G}_{B}(\varepsilon\omega_0) &=& \frac{(u^{0})^{-1}}{4\pi}
\,\sum_{l,m}\Biggl[{\cal P}^{+}_{(\varepsilon\widetilde{\omega}_0 + m\Omega) l m}(r, \theta)
\,\theta(\varepsilon\widetilde{\omega}_0 + m\Omega)
+ {\cal P}^{-}_{(\varepsilon\widetilde{\omega}_0 + m\Omega) l m}(r, \theta)\,
\theta[\varepsilon \widetilde{\omega}_0 + m(\Omega-\Omega_{H})] \Biggr].
\eea
In this way one gets
\beq
\widehat{{\cal A}}_{B}(\omega_0) = \frac{{\cal G}_{B}(\omega_0) - {\cal G}_{B}(-\omega_0)}
{{\cal G}_{B}(\omega_0) + {\cal G}_{B}(-\omega_0)}
= \sum_{l,m} \widehat{{\cal B}}_{lm}(\omega_0)
\eeq
where
\bea
\widehat{{\cal B}}_{lm}(\omega_0) = \frac{\widehat{{\cal W}}_{B}^{lm\,-}(\omega_0)}
{\sum_{l,m}\widehat{{\cal W}}_{B}^{lm\,+}(\omega_0)}
\eea
with
\bea
\widehat{{\cal W}}_{B}^{lm\,\pm}(\omega_0) &=& 
{\cal P}^{+}_{(\widetilde{\omega}_0 + m\Omega) l m}(r, \theta)
\,\theta(\widetilde{\omega}_0 + m\Omega)
+ {\cal P}^{-}_{(\widetilde{\omega}_0 + m\Omega) l m}(r, \theta)\,
\theta[\widetilde{\omega}_0 + m(\Omega-\Omega_{H})]
\nn\\
&\pm&\, {\cal P}^{+}_{(-\widetilde{\omega}_0 + m\Omega) l m}(r, \theta)
\,\theta(-\widetilde{\omega}_0 + m\Omega)
\pm {\cal P}^{-}_{(-\widetilde{\omega}_0 + m\Omega) l m}(r, \theta)\,
\theta[-\widetilde{\omega}_0 + m(\Omega-\Omega_{H})].
\eea
(We have changed slightly our notation so that the results to be discussed now can be readily distinguished from those obtained in the previous section.) Note that the concurrence~(\ref{sep}) cannot reach its maximum value for a finite $r$ due to the Unruh-Starobinskii radiation as well as the existence of a finite angular velocity for the qubits. Let us see how this scenario changes when one considers the regions ${\cal J}^{-}$ and ${\cal H}^{-}$. For the former, since ${\cal P}^{-}$ vanishes in this region and 
$\Omega \to 0$ when $r \to \infty$, one finds that $\widehat{{\cal A}}_{B}(\omega_0) \to 1$ at ${\cal J}^{-}$, reproducing the result of the previous section. On the other hand, near ${\cal H}^{-}$ it is 
${\cal P}^{+}$ that vanishes and  $\Omega \to \Omega_{H}$ close to the horizon. As a result one also finds that $\widehat{{\cal A}}_{B}(\omega_0) \to 1$ at ${\cal H}^{-}$. So we conclude that at ${\cal J}^{-}$ and ${\cal H}^{-}$ the concurrence reaches its maximum possible value in the present scenario ($1/2$) and entanglement between the stationary qubits is accordingly stronger in such regions. Nevertheless, we emphasize that any conclusions regarding the Boulware vacuum state near the event horizon should be understood with due care since such a state is known to be pathological at the event horizon~\cite{candelas2}. In turn, for the complete positivity to hold one must require that $\widehat{{\cal A}}_{B}(\omega_0) \leq 1$, or ${\cal G}_{B}(-\omega_0) \geq 0$, which is always satisfied. This in turn implies that the Markovian limit of the time evolution of the qubit system can be consistently considered in this case.

\subsection{The Unruh vacuum state}

Now let us discuss the case of a gravitational collapse of a rotating massive body. The Fourier transform of the Wightman function reads
\bea
{\cal G}_{U}(\varepsilon\omega_0) &=& \frac{1}{8\pi^2}\sum_{l,m}\,
\,\int_{-\infty}^{\infty}du\,\Biggl\{\int_{0}^{\infty}\,d\omega\,{\cal P}^{+}_{\omega l m}(r, \theta)\, 
\left[\cos \Bigl((\widetilde{w}-\widetilde{m})u\Bigr) -i\sin \Bigl((\widetilde{w}-\widetilde{m})u\Bigr)\right]
\nn\\
&+&\, \int_{0}^{\infty}\,d\bar{\omega}{\cal P}^{-}_{\omega l m}(r, \theta)\,
\left[ \coth\left( \frac{\pi\bar{\omega}}{\kappa_{+}} \right)\cos \Bigl((\widetilde{w}-\widetilde{m})u\Bigr) 
-i\sin \Bigl((\widetilde{w}-\widetilde{m})u\Bigr)\right] \Biggr\}\,e^{i \varepsilon \omega_0 u}.
\eea
Solving the time integrals, one gets
\bea
{\cal G}_{U}(\varepsilon\omega_0) &=& \frac{(u^{0})^{-1}}{4\pi}
\,\sum_{l,m}\Biggl[{\cal P}^{+}_{(\varepsilon\widetilde{\omega}_0 + m\Omega) l m}(r, \theta)
\,\theta(\varepsilon\widetilde{\omega}_0 + m\Omega)
\nn\\
&+&\, \left(1+ \frac{1}{e^{ \frac{2\pi}{\kappa_{+}}[\varepsilon \widetilde{\omega}_0 + m(\Omega-\Omega_{H})]}-1}\right){\cal P}^{-}_{(\varepsilon\widetilde{\omega}_0 + m\Omega) l m}(r, \theta)\,
\theta[\varepsilon\widetilde{\omega}_0 + m(\Omega-\Omega_{H})] 
\nn\\
&+&\,\frac{{\cal P}^{-}_{(-\varepsilon\widetilde{\omega}_0 + m\Omega) l m}(r, \theta)}{e^{ \frac{2\pi}{\kappa_{+}}[-\varepsilon \widetilde{\omega}_0 + m(\Omega-\Omega_{H})]}-1}\,
\theta[-\varepsilon\widetilde{\omega}_0 + m(\Omega-\Omega_{H})]\Biggr].
\eea
In this way one finds
\beq
\widehat{{\cal A}}_{U}(\omega_0) = \frac{{\cal G}_{U}(\omega_0) - {\cal G}_{U}(-\omega_0)}
{{\cal G}_{U}(\omega_0) + {\cal G}_{U}(-\omega_0)}
= \sum_{l,m} \widehat{{\cal U}}_{lm}(\omega_0)
\eeq
where
\bea
\widehat{{\cal U}}_{lm}(\omega_0) = \frac{\widehat{{\cal W}}_{U}^{lm\,-}(\omega_0)}
{\sum_{l,m}\widehat{{\cal W}}_{U}^{lm\,+}(\omega_0)}
\eea
with
\bea
\widehat{{\cal W}}_{U}^{lm\,\pm}(\omega_0) &=& 
{\cal P}^{+}_{(\widetilde{\omega}_0 + m\Omega) l m}(r, \theta)
\,\theta(\widetilde{\omega}_0 + m\Omega)
\nn\\
&+&\, \left(1+ \frac{1}{e^{ \frac{2\pi}{\kappa_{+}}[\widetilde{\omega}_0 + m(\Omega-\Omega_{H})]}-1}\right){\cal P}^{-}_{(\widetilde{\omega}_0 + m\Omega) l m}(r, \theta)\,
\theta[\widetilde{\omega}_0 + m(\Omega-\Omega_{H})] 
\nn\\
&+&\,\frac{{\cal P}^{-}_{(-\widetilde{\omega}_0 + m\Omega) l m}(r, \theta)}{e^{ \frac{2\pi}{\kappa_{+}}[-\widetilde{\omega}_0 + m(\Omega-\Omega_{H})]}-1}\,
\theta[-\widetilde{\omega}_0 + m(\Omega-\Omega_{H})]
\nn\\
&\pm&{\cal P}^{+}_{(-\widetilde{\omega}_0 + m\Omega) l m}(r, \theta)
\,\theta(-\widetilde{\omega}_0 + m\Omega)
\nn\\
&\pm&\, \left(1+ \frac{1}{e^{ \frac{2\pi}{\kappa_{+}}[-\widetilde{\omega}_0 + m(\Omega-\Omega_{H})]}-1}\right){\cal P}^{-}_{(-\widetilde{\omega}_0 + m\Omega) l m}(r, \theta)\,
\theta[-\widetilde{\omega}_0 + m(\Omega-\Omega_{H})] 
\nn\\
&\pm&\,\frac{{\cal P}^{-}_{(\widetilde{\omega}_0 + m\Omega) l m}(r, \theta)}{e^{ \frac{2\pi}{\kappa_{+}}
[\widetilde{\omega}_0 + m(\Omega-\Omega_{H})]}-1}\,
\theta[\widetilde{\omega}_0 + m(\Omega-\Omega_{H})].
\eea
Again superradiance and the finite angular velocity for the qubits prevents the concurrence~(\ref{sep}) from reaching its maximum value for a finite $r$; the difference from the previous case is the presence of the thermal radiation, signaling that the Hawking effect also plays a part in this action.

Let us analyze the asymptotic entanglement within the regions ${\cal J}^{-}$ and ${\cal H}^{-}$. At ${\cal J}^{-}$, due to the backscattering off the spacetime curvature and $\Omega \to 0$,  one finds that the concurrence grows larger, attaining its maximum value of $1/2$. On the other hand, close to ${\cal H}^{-}$,  $\Omega \to \Omega_{H}$ and ${\cal P}^{+}$ vanishes; one gets
\beq
\widehat{{\cal W}}_{U}^{lm\,\pm}(\omega_0) \approx 
 \left(1+ \frac{1}{e^{ \frac{2\pi\widetilde{\omega}_0}{\kappa_{+}}}-1}\right){\cal P}^{-}_{(\widetilde{\omega}_0 + m\Omega_{H}) l m}(r, \theta)\,
\pm
\frac{{\cal P}^{-}_{(\widetilde{\omega}_0 + m\Omega_{H}) l m}(r, \theta)}{e^{ \frac{2\pi\widetilde{\omega}_0}{\kappa_{+}}}-1}
\eeq
where the asymptotic behavior of ${\cal P}^{-}$ at can be read off from Eq.~(\ref{pasymp2}). So inside the ergoregion the Hawking radiation does not allow the asymptotic entanglement to reach its maximum possible value.

Let us investigate the  complete positivity for the evolution of the reduced system. This condition is warranted as long as ${\cal G}_{U}(-\omega_0) \geq 0$, a requirement which one can easily see to hold in the present case. As a result, one may safely assert that the Markovian approximation is warranted.

\subsection{The Candelas-Chrzanowski-Howard vacuum state}

Now we turn to the considerations associated with the candidates of a Hartle-Hawking vacuum states. We begin with the Candelas-Chrzanowski-Howard vacuum state. The Fourier transform of the Wightman function is given by
\bea
{\cal G}_{CCH}(\varepsilon\omega_0) &=& \frac{1}{8\pi^2}\sum_{l,m}\,
\,\int_{-\infty}^{\infty}du\,\Biggl\{\int_{0}^{\infty}\,d\omega\,{\cal P}^{+}_{\omega l m}(r, \theta)\, 
\left[\coth\left( \frac{\pi\omega}{\kappa_{+}} \right)\cos \Bigl((\widetilde{w}-\widetilde{m})u\Bigr) 
-i\sin \Bigl((\widetilde{w}-\widetilde{m})u\Bigr)\right]\nn\\
&+&\, \int_{0}^{\infty}\,d\bar{\omega}{\cal P}^{-}_{\omega l m}(r, \theta)\,
\left[ \coth\left( \frac{\pi\bar{\omega}}{\kappa_{+}} \right)\cos \Bigl((\widetilde{w}-\widetilde{m})u\Bigr) 
-i\sin \Bigl((\widetilde{w}-\widetilde{m})u\Bigr)\right] \Biggr\}\,e^{i \varepsilon \omega_0 u},
\eea
Solving the time integrals, one gets
\bea
{\cal G}_{CCH}(\varepsilon\omega_0) &=& \frac{(u^{0})^{-1}}{4\pi}
\,\sum_{l,m}\Biggl[\left(1+ \frac{1}{e^{ \frac{2\pi}{\kappa_{+}}(\varepsilon \widetilde{\omega}_0 
+ m\Omega)}-1}\right)
{\cal P}^{+}_{(\varepsilon\widetilde{\omega}_0 + m\Omega) l m}(r, \theta)
\,\theta(\varepsilon\widetilde{\omega}_0 + m\Omega)
\nn\\
&+&\,\frac{{\cal P}^{+}_{(-\varepsilon\widetilde{\omega}_0 + m\Omega) l m}(r, \theta)}{e^{ \frac{2\pi}{\kappa_{+}}(-\varepsilon \widetilde{\omega}_0 + m\Omega)}-1}
\,\theta(-\varepsilon\widetilde{\omega}_0 + m\Omega)
\nn\\
&+&\, \left(1+ \frac{1}{e^{ \frac{2\pi}{\kappa_{+}}[\varepsilon \widetilde{\omega}_0 + m(\Omega-\Omega_{H})]}-1}\right){\cal P}^{-}_{(\varepsilon\widetilde{\omega}_0 + m\Omega) l m}(r, \theta)\,
\theta[\varepsilon\widetilde{\omega}_0 + m(\Omega-\Omega_{H})] 
\nn\\
&+&\,\frac{{\cal P}^{-}_{(-\varepsilon\widetilde{\omega}_0 + m\Omega) l m}(r, \theta)}{e^{ \frac{2\pi}{\kappa_{+}}[-\varepsilon \widetilde{\omega}_0 + m(\Omega-\Omega_{H})]}-1}\,
\theta[-\varepsilon\widetilde{\omega}_0 + m(\Omega-\Omega_{H})]\Biggr].
\eea
In this way one gets
\beq
\widehat{{\cal A}}_{CCH}(\omega_0) = \frac{{\cal G}_{CCH}(\omega_0) - {\cal G}_{CCH}(-\omega_0)}
{{\cal G}_{CCH}(\omega_0) + {\cal G}_{CCH}(-\omega_0)}
= \sum_{l,m} \widehat{{\cal F}}_{lm}(\omega_0)
\eeq
where
\bea
\widehat{{\cal F}}_{lm}(\omega_0) = \frac{\widehat{{\cal W}}_{CCH}^{lm\,-}(\omega_0)}
{\sum_{l,m}\widehat{{\cal W}}_{CCH}^{lm\,+}(\omega_0)}
\eea
with
\bea
\widehat{{\cal W}}_{CCH}^{lm\,\pm}(\omega_0) &=& 
\left(1+ \frac{1}{e^{ \frac{2\pi}{\kappa_{+}}(\widetilde{\omega}_0 + m\Omega)}-1}\right) 
{\cal P}^{+}_{(\widetilde{\omega}_0 + m\Omega) l m}(r, \theta)
\,\theta(\widetilde{\omega}_0 + m\Omega)
\nn\\
&+&\,\frac{{\cal P}^{+}_{(-\widetilde{\omega}_0 + m\Omega) l m}(r, \theta)}
{e^{ \frac{2\pi}{\kappa_{+}}(-\widetilde{\omega}_0 + m\Omega)}-1}
\,\theta(-\widetilde{\omega}_0 + m\Omega)
\nn\\
&+&\, \left(1+ \frac{1}{e^{ \frac{2\pi}{\kappa_{+}}[\widetilde{\omega}_0 + m(\Omega-\Omega_{H})]}-1}\right){\cal P}^{-}_{(\widetilde{\omega}_0 + m\Omega) l m}(r, \theta)\,
\theta[\widetilde{\omega}_0 + m(\Omega-\Omega_{H})] 
\nn\\
&+&\,\frac{{\cal P}^{-}_{(-\widetilde{\omega}_0 + m\Omega) l m}(r, \theta)}
{e^{ \frac{2\pi}{\kappa_{+}}[-\widetilde{\omega}_0 + m(\Omega-\Omega_{H})]}-1}\,
\theta[-\widetilde{\omega}_0 + m(\Omega-\Omega_{H})]
\nn\\
&\pm&\,
\left(1+ \frac{1}{e^{ \frac{2\pi}{\kappa_{+}}(-\widetilde{\omega}_0 + m\Omega)}-1}\right)
{\cal P}^{+}_{(-\widetilde{\omega}_0 + m\Omega) l m}(r, \theta)
\,\theta(-\widetilde{\omega}_0 + m\Omega)
\nn\\
&\pm&\,\frac{{\cal P}^{+}_{(\widetilde{\omega}_0 + m\Omega) l m}(r, \theta)}
{e^{ \frac{2\pi}{\kappa_{+}}(\widetilde{\omega}_0 + m\Omega)}-1}
\,\theta(\widetilde{\omega}_0 + m\Omega)
\nn\\
&\pm&\, \left(1+ \frac{1}{e^{ \frac{2\pi}{\kappa_{+}}[-\widetilde{\omega}_0 + m(\Omega-\Omega_{H})]}-1}\right){\cal P}^{-}_{(-\widetilde{\omega}_0 + m\Omega) l m}(r, \theta)\,
\theta[-\widetilde{\omega}_0 + m(\Omega-\Omega_{H})] 
\nn\\
&\pm&\,\frac{{\cal P}^{-}_{(\widetilde{\omega}_0 + m\Omega) l m}(r, \theta)}
{e^{ \frac{2\pi}{\kappa_{+}}[\widetilde{\omega}_0 + m(\Omega-\Omega_{H})]}-1}\,
\theta[\widetilde{\omega}_0 + m(\Omega-\Omega_{H})].
\eea
Again notice the presence of thermal factors in all contributions above. In addition, as above the presence of superradiant modes as well as a finite $\Omega$ prevent a robust entanglement between the qubits, even at the asymptotic regions. 

Let us analyze the entanglement at the asymptotic regions. At ${\cal J}^{-}$ one has
\beq
\widehat{{\cal W}}_{CCH}^{lm\,\pm}(\omega_0) \approx 
\left(1+ \frac{1}{e^{ \frac{2\pi\widetilde{\omega}_0}{\kappa_{+}}}-1}\right) 
{\cal P}^{+}_{\widetilde{\omega}_0 l m}(r, \theta)
\pm \frac{{\cal P}^{+}_{\widetilde{\omega}_0 l m}(r, \theta)}
{e^{ \frac{2\pi\widetilde{\omega}_0}{\kappa_{+}}}-1}.
\eeq
where the asymptotic behavior of ${\cal P}^{+}$ is given by Eq.~(\ref{pasymp1}). In contrast to the previous cases, the concurrence does not attain its maximum at ${\cal J}^{-}$. This is due to the thermal nature of the Candelas-Chrzanowski-Howard vacuum state.  In turn, close to ${\cal H}^{-}$ one gets
\beq
\widehat{{\cal W}}_{CCH}^{lm\,\pm}(\omega_0) \approx 
 \left(1+ \frac{1}{e^{ \frac{2\pi\widetilde{\omega}_0}{\kappa_{+}}}-1}\right)
 {\cal P}^{-}_{(\widetilde{\omega}_0 + m\Omega_{H}) l m}(r, \theta)\,
\pm \frac{{\cal P}^{-}_{(\widetilde{\omega}_0 + m\Omega_{H}) l m}(r, \theta)}
{e^{ \frac{2\pi\widetilde{\omega}_0}{\kappa_{+}}}-1}
\eeq
the asymptotic behavior of ${\cal P}^{-}$ being given by Eq.~(\ref{pasymp2}). Hence in the asymptotic regions the interplay between superradiance and the Hawking effect prevent the asymptotic entanglement to achieve its maximum. On the other hand, the complete positivity for the evolution of the subsystem of the qubits is ensured when ${\cal G}_{CCH}(-\omega_0) \geq 0$, which is clearly satisfied. This also means that the Markovian approximation is well defined in this case.

\subsection{The Frolov-Thorne vacuum state}

To conclude our discussions, we finally turn our attentions to the Frolov-Thorne vacuum state. The Fourier transform of the Wightman function reads
\bea
{\cal G}_{FT}(\varepsilon\omega_0) &=& \frac{1}{8\pi^2}\sum_{l,m}\,
\,\int_{-\infty}^{\infty}du\,\Biggl[\int_{0}^{\infty}\,d\omega\,\coth\left( \frac{\pi\bar{\omega}}{\kappa_{+}} \right)
{\cal P}^{+}_{\omega l m}(r, \theta)\, 
e^{-i (\widetilde{w}-\widetilde{m}) u} 
\nn\\
&+&\, \int_{0}^{\infty}\,d\bar{\omega} \coth\left( \frac{\pi\bar{\omega}}{\kappa_{+}} \right)
{\cal P}^{-}_{\omega l m}(r, \theta)\,
e^{-i (\widetilde{w}-\widetilde{m}) u} \Biggr]\,e^{i \varepsilon \omega_0 u},
\eea
Solving the time integrals, one gets
\bea
{\cal G}_{FT}(\varepsilon\omega_0) &=& \frac{(u^{0})^{-1}}{4\pi}
\,\sum_{l,m}
\Biggl[\left(1+ \frac{1}{e^{ \frac{2\pi}{\kappa_{+}}[\varepsilon \widetilde{\omega}_0 
+ m(\Omega-\Omega_{H})]}-1}\right)
{\cal P}^{+}_{(\varepsilon\widetilde{\omega}_0 + m\Omega) l m}(r, \theta)
\,\theta(\varepsilon\widetilde{\omega}_0 + m\Omega)
\nn\\
&+&\,\frac{{\cal P}^{+}_{(-\varepsilon\widetilde{\omega}_0 + m\Omega) l m}(r, \theta)}
{e^{ \frac{2\pi}{\kappa_{+}}[-\varepsilon \widetilde{\omega}_0 + m(\Omega-\Omega_{H})]}-1}
\,\theta(-\varepsilon\widetilde{\omega}_0 + m\Omega)
\nn\\
&+&\, \left(1+ \frac{1}{e^{ \frac{2\pi}{\kappa_{+}}[\varepsilon \widetilde{\omega}_0 + m(\Omega-\Omega_{H})]}-1}\right){\cal P}^{-}_{(\varepsilon\widetilde{\omega}_0 + m\Omega) l m}(r, \theta)\,
\theta[\varepsilon\widetilde{\omega}_0 + m(\Omega-\Omega_{H})] 
\nn\\
&+&\,\frac{{\cal P}^{-}_{(-\varepsilon\widetilde{\omega}_0 + m\Omega) l m}(r, \theta)}{e^{ \frac{2\pi}{\kappa_{+}}[-\varepsilon \widetilde{\omega}_0 + m(\Omega-\Omega_{H})]}-1}\,
\theta[-\varepsilon\widetilde{\omega}_0 + m(\Omega-\Omega_{H})]\Biggr].
\eea
In this way one finds
\beq
\widehat{{\cal A}}_{FT}(\omega_0) = \frac{{\cal G}_{FT}(\omega_0) - {\cal G}_{FT}(-\omega_0)}
{{\cal G}_{FT}(\omega_0) + {\cal G}_{FT}(-\omega_0)}
= \sum_{l,m} \widehat{{\cal X}}_{lm}(\omega_0)
\eeq
where
\bea
\widehat{{\cal X}}_{lm}(\omega_0) = \frac{\widehat{{\cal W}}_{FT}^{lm\,-}(\omega_0)}
{\sum_{l,m}\widehat{{\cal W}}_{FT}^{lm\,+}(\omega_0)}
\eea
with
\bea
\widehat{{\cal W}}_{FT}^{lm\,\pm}(\omega_0) &=& \left(1+ \frac{1}{e^{ \frac{2\pi}{\kappa_{+}}[\widetilde{\omega}_0 + m(\Omega-\Omega_{H})]}-1}\right) 
{\cal P}^{+}_{(\widetilde{\omega}_0 + m\Omega) l m}(r, \theta)
\,\theta(\widetilde{\omega}_0 + m\Omega)
\nn\\
&+&\,\frac{{\cal P}^{+}_{(-\widetilde{\omega}_0 + m\Omega) l m}(r, \theta)}
{e^{ \frac{2\pi}{\kappa_{+}}[-\widetilde{\omega}_0 + m(\Omega-\Omega_{H})]}-1}
\,\theta(-\widetilde{\omega}_0 + m\Omega)
\nn\\
&+&\, \left(1+ \frac{1}{e^{ \frac{2\pi}{\kappa_{+}}[\widetilde{\omega}_0 + m(\Omega-\Omega_{H})]}-1}\right){\cal P}^{-}_{(\widetilde{\omega}_0 + m\Omega) l m}(r, \theta)\,
\theta[\widetilde{\omega}_0 + m(\Omega-\Omega_{H})] 
\nn\\
&+&\,\frac{{\cal P}^{-}_{(-\widetilde{\omega}_0 + m\Omega) l m}(r, \theta)}
{e^{ \frac{2\pi}{\kappa_{+}}[-\widetilde{\omega}_0 + m(\Omega-\Omega_{H})]}-1}\,
\theta[-\widetilde{\omega}_0 + m(\Omega-\Omega_{H})]
\nn\\
&\pm&\,
\left(1+ \frac{1}{e^{ \frac{2\pi}{\kappa_{+}}[-\widetilde{\omega}_0 + m(\Omega-\Omega_{H})]}-1}\right)
{\cal P}^{+}_{(-\widetilde{\omega}_0 + m\Omega) l m}(r, \theta)
\,\theta(-\widetilde{\omega}_0 + m\Omega)
\nn\\
&\pm&\,\frac{{\cal P}^{+}_{(\widetilde{\omega}_0 + m\Omega) l m}(r, \theta)}
{e^{ \frac{2\pi}{\kappa_{+}}[\widetilde{\omega}_0 + m(\Omega-\Omega_{H})]}-1}
\,\theta(\widetilde{\omega}_0 + m\Omega)
\nn\\
&\pm&\, \left(1+ \frac{1}{e^{ \frac{2\pi}{\kappa_{+}}[-\widetilde{\omega}_0 + m(\Omega-\Omega_{H})]}-1}\right){\cal P}^{-}_{(-\widetilde{\omega}_0 + m\Omega) l m}(r, \theta)\,
\theta[-\widetilde{\omega}_0 + m(\Omega-\Omega_{H})] 
\nn\\
&\pm&\,\frac{{\cal P}^{-}_{(\widetilde{\omega}_0 + m\Omega) l m}(r, \theta)}
{e^{ \frac{2\pi}{\kappa_{+}}[\widetilde{\omega}_0 + m(\Omega-\Omega_{H})]}-1}\,
\theta[\widetilde{\omega}_0 + m(\Omega-\Omega_{H})].
\eea
In this case there are some subtleties regarding the time evolution of the qubit subsystem. Before we analyze this issue, let us discuss the formation of entanglement at the asymptotic regions. At ${\cal J}^{-}$ one finds
\beq
\widehat{{\cal W}}_{FT}^{lm\,\pm}(\omega_0) \approx 
\left(1+ \frac{1}{e^{ \frac{2\pi}{\kappa_{+}}(\widetilde{\omega}_0 - m\Omega_{H})}-1}\right) 
{\cal P}^{+}_{\widetilde{\omega}_0 l m}(r, \theta)
\pm\frac{{\cal P}^{+}_{\widetilde{\omega}_0 l m}(r, \theta)}
{e^{ \frac{2\pi}{\kappa_{+}}(\widetilde{\omega}_0 - m\Omega_{H})}-1}
\eeq
and the asymptotic behavior of ${\cal P}^{+}$ is given by Eq.~(\ref{pasymp1}). In turn, close to ${\cal H}^{-}$ one gets
\bea
\widehat{{\cal W}}_{FT}^{lm\,\pm}(\omega_0) &\approx& 
 \left(1+ \frac{1}{e^{ \frac{2\pi\widetilde{\omega}_0}{\kappa_{+}}}-1}\right)
 {\cal P}^{-}_{(\widetilde{\omega}_0 + m\Omega_{H}) l m}(r, \theta)\,
\pm \frac{{\cal P}^{-}_{(\widetilde{\omega}_0 + m\Omega_{H}) l m}(r, \theta)}
{e^{ \frac{2\pi\widetilde{\omega}_0}{\kappa_{+}}}-1}\,
\eea
and the asymptotic behavior of ${\cal P}^{-}$ is given by Eq.~(\ref{pasymp2}). So near ${\cal H}^{-}$ we found the same asymptotic behavior for the Unruh, Frolov-Thorne and Candelas-Chrzanowski-Howard vacuum states. The difference lies when the qubits depart from the black hole. In order to understand the consequences of such distinct results, let us look closely at the requirement of complete positivity given by  ${\cal G}_{FT}(-\omega_0) \geq 0$. From the results derived above, one can show that this is equivalent to the inequality
\bea
&&\sum_{l,m} \left\{
\left(1+ \frac{1}{e^{ \frac{2\pi}{\kappa_{+}}[-\widetilde{\omega}_0 + m(\Omega-\Omega_{H})]}-1}\right)
{\cal P}^{+}_{(-\widetilde{\omega}_0 + m\Omega) l m}(r, \theta)
\,\theta(-\widetilde{\omega}_0 + m\Omega)
+ \frac{{\cal P}^{+}_{(\widetilde{\omega}_0 + m\Omega) l m}(r, \theta)}
{e^{ \frac{2\pi}{\kappa_{+}}[\widetilde{\omega}_0 + m(\Omega-\Omega_{H})]}-1}
\,\theta(\widetilde{\omega}_0 + m\Omega)
\right.
\nn\\
&&+\,\left. \left(1+ \frac{1}{e^{ \frac{2\pi}{\kappa_{+}}[-\widetilde{\omega}_0 + m(\Omega-\Omega_{H})]}-1}\right){\cal P}^{-}_{(-\widetilde{\omega}_0 + m\Omega) l m}(r, \theta)\,
\theta[-\widetilde{\omega}_0 + m(\Omega-\Omega_{H})] 
\right.
\nn\\
&&+\,\left. \frac{{\cal P}^{-}_{(\widetilde{\omega}_0 + m\Omega) l m}(r, \theta)}
{e^{ \frac{2\pi}{\kappa_{+}}[\widetilde{\omega}_0 + m(\Omega-\Omega_{H})]}-1}\,
\theta[\widetilde{\omega}_0 + m(\Omega-\Omega_{H})]\right\} \geq 0.
\label{pos-FT}
\eea
At ${\cal J}^{-}$ one has that $\Omega \to 0$ and ${\cal P}^{-}$ vanishes, and the only contribution comes from the second term. This in turn could be negative whether the energy gap obeys the superradiant condition $\widetilde{\omega}_0 < m\Omega_{H}$ for some values of $m$. Hence complete positivity is not ensured in this region and the initial state could be mapped out of the physical state-space by the time evolution. As discussed previously, this is a consequence of the Unruh-Starobinskii effect. In this situation, one has a negative spontaneous excitation rate. As in the case of the static atoms, the expedient of Markovian approximation might lead to inconsistent results whenever one considers the quantum field prepared in the Frolov-Thorne vacuum state.

\section{Conclusions and Perspectives}
\label{conclude}

The purpose of this work was to search for an interpretation of the interplay between the Hawking and the Unruh-Starobinskii effects in an open-system framework. Here we probed the generation of entanglement between qubits in a rotating black-hole background. For static qubits, we found that asymptotic entanglement could be maximized as one approaches the past null infinity depending on the choice of the vacuum state for the quantum fields. This outcome appears to indicate that no evolution from a separable state could lead to an asymptotic entangled state with the maximum value $1/2$ for the concurrence for the cases considered in this paper; however, note that the attainment of the maximum asymptotic entanglement is possible for the future Boulware vacuum state at ${\cal J}^{+}$. This can be seen from a direct analysis of the asymptotic properties of relevant field modes~\cite{ote}. On the other hand, as one approaches the black hole the entanglement is weakened, due to the influences of Unruh-Starobinskii and Hawking effects. Indeed, for the Candelas-Chrzanowski-Howard we notice that the asymptotic state of the static qubits does not appear to be entangled when both are close to the ergosphere.

In the ZAMOs perspective, the aforementioned physical picture undergoes some modifications. In this case a peculiar pattern was observed for the Boulware vacuum state of the field, namely a stronger asymptotic entanglement near the event horizon and at spatial infinity. This suggests the presence of a kind of entanglement revival for stationary qubits with zero angular momentum in this situation. However, as previously observed, such conclusions should be perceived with adequate caution, since the Boulware vacuum state is pathological at the horizon in the sense that the renormalized expectation value of the energy-momentum tensor diverges as $r \to r_{+}$. We also notice that the backscattering of the Hawking radiation due to the curvature for the Unruh vacuum facilitates the generation of asymptotic entanglement as one gets farther away from the black hole. This is not observed for the Candelas-Chrzanowski-Howard state.

In almost all cases analyzed we have not verified a violation of complete positivity for the time evolution of the atomic subsystem. The only exception lies within the Frolov-Thorne vacuum state, whose results must be taken with care. As shown above, at spatial infinity, the time evolution of the qubit subsystem could cease to maintain complete positivity in this case. A possible way to circumvent this nuisance would be to represent the Kossakowski matrix ${\cal K}$ as the sum of two matrices, one with positive eigenvalues and the other with the negative ones. By extracting an overall minus sign, the Kossakowski matrix can be pictured as the difference of two positive definite matrices, ${\cal K} = {\cal K}_1 - {\cal K}_2$. The drawback in this argument is that the feature of complete positivity determines the structure of dynamical maps by ruling out the presence of a term such as ${\cal K}_2$, whereas the issue of positivity is not completely settled down. Indeed, no general prescriptions on the matrices ${\cal K}_1,{\cal K}_2$ are known in order to assure the positivity of ${\cal K}$~\cite{ben3}.

In any case, despite the matter raised by the Frolov-Thorne vacuum state, the overall physical picture that emerges in our studies is the following. Appreciable differences between the results found in Kerr spacetime and Schwarzschild spacetimes can be drawn. In the former, superradiance affects in a nontrivial way the dynamics of open quantum systems in a Kerr background; it can be an obstruction for a higher degree of entanglement, as well as a barrier to the complete positivity. The requirement that supports this view is given whenever the energy gap obeys a kind of superradiant condition. One also finds that, for a quantum field prepared in the Boulware vacuum state, the concurrence reaches its maximum only at the asymptotic regions. In turn, for stationary qubits the Unruh and the Candelas-Chrzanowski-Howard states present a similar deportment inside the ergoregion; as one gets closer to the event horizon both states display the same asymptotic behavior.  By contrast, the Frolov-Thorne vacuum state may disrupt the complete positivity of dynamical maps as discussed. As remarked above, this could be definitely interpreted as a possible manifestation of the breakdown of the Markov approximation. The origin of this inconsistency lies, of course, in the presence of superradiant modes which causes the emergence of the Unruh-Starobinskii effect. We highlight that this is only verified for the Frolov-Thorne vacuum state, since it is only in this case that the spontaneous excitation rate is clearly negative (when the energy gap of the qubits satisfy the superradiant condition), which could be interpreted as an evidence of the breakdown of the Markovian approximation. In all other cases analyzed in this work the presence of superradiant modes does not seem to upset the Markovian limit of the evolution of the atomic system, and hence complete positivity is maintained. In this situation, one should resort to a more general analysis in order to avoid a physically inconsistent time-evolution that violates the positivity of the spectrum of initial density matrices. On the other hand, one could also argue that the Frolov-Thorne vacuum state is not an adequate choice of vacuum state (except for qubits located on the axis, where the $m \neq 0$ terms do not contribute), since it has been shown that this state fails to be regular almost everywhere, both on or outside the event horizon~\cite{ote}. Indeed, the existence of the weak-coupling limit (on which the Markovian approximation is based) is connected with certain properties that the Wightman functions must satisfy~\cite{edavies1}. It could be the case that the Frolov-Thorne vacuum state does not fulfill all such conditions. A detailed study of this aspect would be most welcome, but it is outside of the scope of the present article.

This work suggests a possible path to assemble concepts from black-hole superradiance, relativistic jets and quantum information theory in the light of open quantum systems. To a large extent researches on quantum entanglement in a black-hole background are relevant due to the problem of black-hole information loss. The general conception is that, by establishing certain relationships between the equivalence principle and quantum entanglement, a distinguished aspect concerning the black-hole information paradox could be uncovered. 

On the other hand, there are natural extensions of this work concerning primarily the field of relativistic quantum information. For instance, an important phenomenon widely discussed in the recent literature is the extraction of non-classical correlations from the quantum vacuum to qubits (or particle detectors). This is known as entanglement harvesting. The possibility of harvesting entanglement from quantum fields has prompted various works to investigate this effect in several different scenarios~\cite{reznik1,martin2,Reznick:05,Valentini:91,Pozas:15,Steeg:09,martin3}. For a recent discussion concerning entanglement harvesting from black holes, see Ref.~\cite{Henderson:17}. In order to study this phenomenon within our setting, one would have to implement a variety of substantial modifications. Firstly, one would have to consider spatially separated qubits. This implies that each qubit is moving along a different stationary trajectory, each of which parametrized by different proper times. This issue can be dealt with by considering the time evolution with respect to the Boyer-Lindquist coordinate time $t$. This will bring in to each quantity in the master equation an explicit dependence on the labels of each qubit, $a,b$. Hence one would have to work with a more general Kossakowski matrix $K^{(a)(b)}(\omega_0)$. Moreover, an overall oscillatory time-dependent factor multiplying the dissipative contribution would appear due to the usual gravitational redshift effect [besides an overall factor of $(g_{00}(x_{(a)})g_{00}(x_{(b)}))^{1/2}$]. This is not of great concern, since one could, in principle, resort to the secular approximation and argue that the main contribution to the dissipative term would come when this factor is reduced to unity. Nevertheless, one would have to investigate the solution to the master equation at finite time and then check the possible existence of an equilibrium state in order to verify whether this secular approximation can be employed. These are necessary steps in order to avoid any inconsistencies, such as reduced density operators at a finite time with negative eigenvalues. This topic certainly deserves further investigations and future studies will be devoted to this matter. In any case, we believe that studies comprising rotating black holes may contribute to achieve more insights on the quantum structure of the spacetime as well as open questions regarding gravitational collapse.

\section*{Acknowledgements}

The author acknowledges the hospitality of the Institute for Theoretical Physics, University of Cologne, where part of this research was completed. He also thanks L. H. Ford and C. Kiefer for useful discussions. Valuable comments from an anonymous referee are also acknowledged. 

\appendix

\section{Wightman functions of a massless scalar field in Kerr spacetime}
\label{B}

In order to present an explicit expression for the Wightman function in a curved background, one must first define what one means by the ``vacuum state" of the quantum field~\cite{birrel}. In the case of a rotating black hole, some additional features arise as well~\cite{ote}. This is related with the existence of superradiant modes. In the case of the Kerr black hole, one may define two kinds of ``Boulware" vacuum states and two associated ``Unruh" vacuum states. Such states describe different physical situations. In turn, a true Hartle-Hawking vacuum state (which describes a state of thermal equilibrium between the black hole and its surroundings) for the Kerr black hole cannot be stipulated. In the literature we have two proposals for such a state, namely the Candelas-Chrzanowski-Howard vacuum state and Frolov-Thorne vacuum state. For a recent discussion on the physical meaning of each of those proposed vacuum states in the context of atomic radiative processes, see Ref.~\cite{prd17} 

In order to derive the Wightman functions of a massless scalar field associated with the vacuum states discussed above, one must first outline the quantization of the scalar field in Kerr spacetime~\cite{ote,fro3,ford}. The associated wave equation for the field is separable in the Kerr metric and the basis functions reads
\beq
u_{\omega l m} = N_{\omega l m}\,\frac{e^{-i\omega t + i m \phi}}{(r^2 + a^2)^{1/2}}\,S_{\omega l m}(\cos\theta)\,
R_{\omega l m} (r),
\label{scalar-modes}
\eeq
where $N_{\omega l m}$ is a normalization constant, $l$ and $m$ are integers with $|m| \leq l$. The function $S_{\omega l m}(x)$ is a spheroidal harmonic. The radial equation is given by
\beq
\left[\frac{d^2}{dr_{*}^2} - V_{\omega l m}(r)\right]R_{\omega l m} (r) = 0,
\label{radial}
\eeq
where
\bea
V_{\omega l m}(r) &=& - \left(\omega - \frac{ma}{r^2 + a^2}\right)^2 + \lambda_{lm}(a\omega)\,\frac{\Delta}{(r^2 + a^2)^2} 
+ \frac{2(Mr - a^2)\Delta}{(r^2 + a^2)^3} + \frac{3a^2\Delta^2}{(r^2 + a^2)^4}
\eea
with the tortoise coordinate $r_{*}$ defined as usual by
\beq
r_{*} = \int\,dr\,\frac{r^2 + a^2}{\Delta} = r + \frac{1}{2\kappa_{+}}\,\ln|r - r_{+}| + \frac{1}{2\kappa_{-}}\,\ln|r - r_{-}|
\label{tortoise}
\eeq
and the surface gravity on the inner and outer horizons being given by
\beq
\kappa_{\pm} = \frac{r_{\pm} - r_{\mp}}{2(r_{\pm}^{2} + a^2)}.
\label{surface-gravity}
\eeq
On the other hand
\bea
R^{-}_{\omega l m}(r) &\approx& e^{i\bar{\omega}r_{*}} + {\cal R}^{-}_{\omega l m}e^{-i\bar{\omega}r_{*}},\,\,\,r \to r_{+}
\nn\\
R^{-}_{\omega l m}(r) &\approx& {\cal T}^{-}_{\omega l m}e^{i\omega r_{*}},\,\,\,r \to \infty.
\label{asymp1}
\eea
and
\bea
R^{+}_{\omega l m}(r) &\approx& {\cal T}^{+}_{\omega l m}e^{-i\bar{\omega}r_{*}},\,\,\,r \to r_{+}
\nn\\
R^{+}_{\omega l m}(r) &\approx& e^{-i\omega r_{*}} + {\cal R}^{+}_{\omega l m}e^{i\omega r_{*}},\,\,\,r \to \infty.
\label{asymp2}
\eea
with $\bar{\omega} = \omega - m\Omega_{H}$ and we have defined the transmission and reflection coefficients by ${\cal T}^{\pm}$ and ${\cal R}^{\pm}$, respectively. One has $|{\cal R}^{-}|^2 > 1$ for modes with $\bar{\omega} < 0$ and $\omega > 0$. This means that such modes are reflected back to ${\cal H}^{+}$ with an amplitude greater than they had at ${\cal H}^{-}$. This is known as the superradiance phenomenon. This is also verified for ${\cal R}^{+}$ at ${\cal J}^{\pm}$. We employ standard notation: ${\cal J}^{-}$ (${\cal J}^{+}$) is the past (future) null infinity and the region ${\cal H}^{-}$ (${\cal H}^{+}$) is the past (future) event horizon.

The existence of superradiant modes engenders some complications in the definition of positive-frequency modes. One defines states with particular properties along a given Cauchy surface. For instance, consider the Cauchy surface ${\cal J}^{-} \bigcup {\cal H}^{-}$. The (past) mode basis reads ($\omega > 0$):
\bea
u^{in}_{\omega l m} &=& \left[8\pi^2\omega(r^2 + a^2)\right]^{-1/2}\,e^{-i\omega t + i m \phi}\,S_{\omega l m}(\cos\theta)\,
R^{+}_{\omega l m} (r),\,\,\,\bar{\omega} > - m\Omega_{H}
\nn\\ 
u^{up}_{\omega l m} &=& \left[8\pi^2\bar{\omega}(r^2 + a^2)\right]^{-1/2}\,e^{-i\omega t + i m \phi}\,
S_{\omega l m}(\cos\theta)\,R^{-}_{\omega l m} (r),\,\,\,\bar{\omega} > 0
\nn\\
u^{up}_{-\omega l -m} &=& \left[8\pi^2(-\bar{\omega})(r^2 + a^2)\right]^{-1/2}\,e^{i\omega t - i m \phi}\,
S_{\omega l m}(\cos\theta)\,R^{-}_{-\omega l -m} (r),\,\,\,0 >  \bar{\omega} > -m\Omega_{H},
\eea
where we used that $S_{-\omega l -m} = S_{\omega l m}$. Such modes are orthonormal in the sense of the standard inner product~\cite{ote,fro3}. In turn, for the Cauchy surface $\Sigma = {\cal J}^{+} \bigcup {\cal H}^{+}$ the (future) basis is given by
\bea
u^{out}_{\omega l m} &=& \left[8\pi^2\omega(r^2 + a^2)\right]^{-1/2}\,e^{-i\omega t + i m \phi}\,S_{\omega l m}(\cos\theta)\,
R^{+\,*}_{\omega l m} (r),\,\,\,\bar{\omega} > - m\Omega_{H}
\nn\\ 
u^{down}_{\omega l m} &=& \left[8\pi^2\bar{\omega}(r^2 + a^2)\right]^{-1/2}\,e^{-i\omega t + i m \phi}\,
S_{\omega l m}(\cos\theta)\,R^{-\,*}_{\omega l m} (r),\,\,\,\bar{\omega} > 0
\nn\\
u^{down}_{-\omega l -m} &=& \left[8\pi^2|\bar{\omega}|(r^2 + a^2)\right]^{-1/2}\,e^{i\omega t - i m \phi}\,
S_{\omega l m}(\cos\theta)\,R^{-\,*}_{-\omega l -m} (r),\,\,\,0 >  \bar{\omega} > -m\Omega_{H}.
\eea
These modes are also orthonormal in the sense quoted above. The asymptotic forms of all such modes is fully discussed in Ref.~\cite{ote}.

The quantization of a scalar field in the Kerr metric now proceeds within traditional canonical methods. The scalar field $\varphi$ may be expanded in terms of any of the sets of mode functions:
\bea
\varphi(x) &=&\,\sum_{l,m}\left[\int_{0}^{\infty}\,d\omega ({\hat a}^{a}_{\omega l m} u^{a}_{\omega l m} + 
{\hat a}^{a\,\dagger}_{\omega l m} u^{a*}_{\omega l m}) + \int_{0}^{\infty}\,d\bar{\omega} 
({\hat a}^{b}_{\omega l m} u^{b}_{\omega l m} + {\hat a}^{b\,\dagger}_{\omega l m} u^{b*}_{\omega l m}) \right],
\eea
where $a = \textrm{in, out}$ and $b = \textrm{up, down}$. The nonvanishing commutation relations for the creation and annihilation operators are given by
\bea
&&[{\hat a}^{a}_{\omega l m}, {\hat a}^{a\,\dagger}_{\omega' l' m'}] = \delta(\omega - \omega')\,\delta_{ll'}\,\delta_{mm'},\,\,\,
\bar{\omega} > - m\Omega_{H}
\nn\\
&&[{\hat a}^{b}_{\omega l m}, {\hat a}^{b\,\dagger}_{\omega' l' m'}] = \delta(\omega - \omega')\,\delta_{ll'}\,\delta_{mm'},
\,\,\,\bar{\omega} > 0
\nn\\
&&[{\hat a}^{b}_{-\omega l -m}, {\hat a}^{b\,\dagger}_{-\omega' l' -m'}] = \delta(\omega - \omega')\,\delta_{ll'}\,\delta_{mm'},
\,\,\, 0 >  \bar{\omega} > -m\Omega_{H}.
\eea
We define a past Boulware vacuum state by
\bea
&& {\hat a}^{in}_{\omega l m}|B^{-}\rangle = 0,\,\,\,\bar{\omega} > - m\Omega_{H}
\nn\\
&& {\hat a}^{up}_{\omega l m}|B^{-}\rangle = 0,\,\,\,\bar{\omega} > 0
\nn\\
&& {\hat a}^{up}_{-\omega l -m}|B^{-}\rangle = 0,\,\,\, 0 >  \bar{\omega} > -m\Omega_{H},
\eea
which corresponds to absence of particles in ${\cal J}^{-}$ and ${\cal H}^{-}$. The future Boulware vacuum state is defined by
\bea
&& {\hat a}^{out}_{\omega l m}|B^{+}\rangle = 0,\,\,\,\bar{\omega} > - m\Omega_{H}
\nn\\
&& {\hat a}^{down}_{\omega l m}|B^{+}\rangle = 0,\,\,\,\bar{\omega} > 0
\nn\\
&& {\hat a}^{down}_{-\omega l -m}|B^{+}\rangle = 0,\,\,\, 0 >  \bar{\omega} > -m\Omega_{H},
\eea
which is related to absence of particles in the regions ${\cal J}^{+}$ and ${\cal H}^{+}$. The past Unruh vacuum state $|U^{-}\rangle$ is the state which is empty at ${\cal J}^{-}$ but with the up modes thermally populated. An analogous definition holds for the future Unruh vacuum state $|U^{+}\rangle$. It is $|U^{-}\rangle$ that embodies the state arising at late times from the collapse of a star to a black hole. Accordingly, this is the Unruh vacuum state adopted in this investigation. On the other hand, in order to define a thermal state with most of the properties of the Hartle-Hawking state we consider the two traditional proposals: The vacuum state $|CCH\rangle$~\cite{candelas2}, which is constructed by thermalizing the in and up modes with respect to their natural energy, and the vacuum state $|FT\rangle$~\cite{fro3}. The former generates a state which does not respect the simultaneous $t-\phi$ reversal invariance of Kerr spacetime. On the other hand, the latter is formally invariant under simultaneous $t-\phi$ reversal. For a detailed discussion on these three vacuum states, see Ref.~\cite{fro3}. 

Now let us present the Wightman functions associated with the vacuum states just described. One can express such functions with the help of the Hadamard function and the Pauli-Jordan function since
$$
\varphi(x)\varphi(x') = \frac{1}{2}\Bigl(\{\varphi(x),\varphi(x')\} + [\varphi(x),\varphi(x')] \Bigr).
$$
Such functions were presented in Ref.~\cite{prd17}. We reproduce them here for convenience. The Hadamard's elementary function of the Boulware vacuum states is given by
\bea
 \langle B^{\pm} |\{\varphi(x),\varphi(x')\}| B^{\pm} \rangle
&=&\, \sum_{l,m}\left\{\int_{0}^{\infty}\,d\omega \left[u^{a}_{\omega l m}(x)u^{a*}_{\omega l m}(x') 
+  u^{a}_{\omega l m}(x')u^{a*}_{\omega l m}(x)\right]\right\} 
\nn\\
&+&\, \sum_{l,m}\left\{\int_{0}^{\infty}\,d\bar{\omega} \left[u^{b}_{\omega l m}(x) u^{b*}_{\omega l m}(x')
+u^{b}_{\omega l m}(x') u^{b*}_{\omega l m}(x)\right] \right\},
\label{hada-boul}
\eea
whereas the associated Pauli-Jordan function reads
\bea
\langle B^{\pm} |[\varphi(x),\varphi(x')]| B^{\pm} \rangle
&=&\, \sum_{l,m}\left\{\int_{0}^{\infty}\,d\omega \left[u^{a}_{\omega l m}(x)u^{a*}_{\omega l m}(x') 
-  u^{a}_{\omega l m}(x')u^{a*}_{\omega l m}(x)\right]\right\} 
\nn\\
&+&\, \sum_{l,m}\left\{\int_{0}^{\infty}\,d\bar{\omega} \left[u^{b}_{\omega l m}(x) u^{b*}_{\omega l m}(x')
- u^{b}_{\omega l m}(x') u^{b*}_{\omega l m}(x)\right] \right\}.
\label{pauli-boul}
\eea
The Hadamard's elementary function of the (past) Unruh vacuum state reads
\bea
\langle U^{-} |\{\varphi(x),\varphi(x')\}| U^{-} \rangle
&=&\, \sum_{l,m}\left\{\int_{0}^{\infty}\,d\omega \left[u^{in}_{\omega l m}(x)u^{in\,*}_{\omega l m}(x') 
+  u^{in}_{\omega l m}(x')u^{in\,*}_{\omega l m}(x)\right]\right\} 
\nn\\
&+&\, \sum_{l,m}\left\{\int_{0}^{\infty}\,d\bar{\omega} \coth\left( \frac{\pi\bar{\omega}}{\kappa_{+}} \right) \left[u^{up}_{\omega l m}(x) u^{up\,*}_{\omega l m}(x')
+u^{up}_{\omega l m}(x') u^{up\,*}_{\omega l m}(x)\right] \right\},
\label{hada-unruh}
\eea
whereas the associated Pauli-Jordan function is given by
\bea
\langle U^{-} |[\varphi(x),\varphi(x')]| U^{-} \rangle
&=&\, \sum_{l,m}\left\{\int_{0}^{\infty}\,d\omega \left[u^{in}_{\omega l m}(x)u^{in\,*}_{\omega l m}(x') 
-  u^{in}_{\omega l m}(x')u^{in\,*}_{\omega l m}(x)\right]\right\} 
\nn\\
&+&\, \sum_{l,m}\left\{\int_{0}^{\infty}\,d\bar{\omega} \left[u^{up}_{\omega l m}(x) u^{up\,*}_{\omega l m}(x')
- u^{up}_{\omega l m}(x') u^{up\,*}_{\omega l m}(x)\right] \right\}.
\label{pauli-unruh}
\eea
In turn, the Hadamard's elementary function of the Candelas-Chrzanowski-Howard vacuum state is given by:
\bea
\langle CCH |\{\varphi(x),\varphi(x')\}| CCH \rangle
&=&\, \sum_{l,m}\left\{\int_{0}^{\infty}\,d\omega\coth\left( \frac{\pi\omega}{\kappa_{+}} \right)
 \left[u^{in}_{\omega l m}(x)u^{in\,*}_{\omega l m}(x') 
+  u^{in}_{\omega l m}(x')u^{in\,*}_{\omega l m}(x)\right]\right\} 
\nn\\
&+&\, \sum_{l,m}\left\{\int_{0}^{\infty}\,d\bar{\omega} \coth\left( \frac{\pi\bar{\omega}}{\kappa_{+}} \right) \left[u^{up}_{\omega l m}(x) u^{up\,*}_{\omega l m}(x')
+u^{up}_{\omega l m}(x') u^{up\,*}_{\omega l m}(x)\right] \right\}.
\label{hada-cCh}
\eea
The associated Pauli-Jordan function reads
\bea
\langle CCH |[\varphi(x),\varphi(x')]| CCH \rangle
&=&\, \sum_{l,m}\left\{\int_{0}^{\infty}\,d\omega \left[u^{in}_{\omega l m}(x)u^{in\,*}_{\omega l m}(x') 
-  u^{in}_{\omega l m}(x')u^{in\,*}_{\omega l m}(x)\right]\right\} 
\nn\\
&+&\, \sum_{l,m}\left\{\int_{0}^{\infty}\,d\bar{\omega} \left[u^{up}_{\omega l m}(x) u^{up\,*}_{\omega l m}(x')
- u^{up}_{\omega l m}(x') u^{up\,*}_{\omega l m}(x)\right] \right\}.
\label{pauli-cCh}
\eea
At last, the Hadamard's elementary function associated with the Frolov-Thorne vacuum state is given by:
\bea
\langle FT |{\hat \eta}\varphi(x) {\hat \eta} \varphi(x') {\hat \eta}| FT \rangle &+&
\langle FT |{\hat \eta}\varphi(x') {\hat \eta} \varphi(x) {\hat \eta}| FT \rangle
\nn\\
&=&\, \sum_{l,m}\left\{\int_{0}^{\infty}\,d\omega\coth\left( \frac{\pi\bar{\omega}}{\kappa_{+}} \right)
 \left[u^{in}_{\omega l m}(x)u^{in\,*}_{\omega l m}(x') 
+  u^{in}_{\omega l m}(x')u^{in\,*}_{\omega l m}(x)\right]\right\} 
\nn\\
&+&\, \sum_{l,m}\left\{\int_{0}^{\infty}\,d\bar{\omega} \coth\left( \frac{\pi\bar{\omega}}{\kappa_{+}} \right) \left[u^{up}_{\omega l m}(x) u^{up\,*}_{\omega l m}(x')
+u^{up}_{\omega l m}(x') u^{up\,*}_{\omega l m}(x)\right] \right\}.
\label{hada-frolov}
\eea
where the number operator ${\hat \eta}^2 = 1$~\cite{fro3}. The Pauli-Jordan function of the Frolov-Thorne vacuum state reads:
\bea
\hspace{-5mm}
\langle FT |{\hat \eta}\varphi(x) {\hat \eta} \varphi(x') {\hat \eta}| FT \rangle -
\langle FT |{\hat \eta}\varphi(x') {\hat \eta} \varphi(x) {\hat \eta}| FT \rangle
&=&\, \sum_{l,m}\left\{\int_{0}^{\infty}\,d\omega \left[u^{in}_{\omega l m}(x)u^{in\,*}_{\omega l m}(x') 
-  u^{in}_{\omega l m}(x')u^{in\,*}_{\omega l m}(x)\right]\right\} 
\nn\\
&+&\, \sum_{l,m}\left\{\int_{0}^{\infty}\,d\bar{\omega} \left[u^{up}_{\omega l m}(x) u^{up\,*}_{\omega l m}(x')
- u^{up}_{\omega l m}(x') u^{up\,*}_{\omega l m}(x)\right] \right\}.
\label{pauli-frolov}
\eea

\end{document}